\documentstyle[bezier,12pt]{article}

\def \R{I\!\!R}

\def \M{{\cal M}}

\def\lan{\mathop{\langle}}
\def\ran{\mathop{\rangle}}

\def\f{{\varphi}}
\def\F{{\cal F}}

\def\eps{\varepsilon}
\def\g{\gamma}
\def\d{\delta}

\date{November 11, 2004}

\title{ DIVERSITY AND RELATIVE ARBITRAGE IN  EQUITY MARKETS}

\vspace{2mm}

\author{ ROBERT FERNHOLZ                  \\\small INTECH,  One Palmer Square    \\\small Princeton, NJ 08542                        \\\small {\tt bob@enhanced.com}
          \and IOANNIS KARATZAS\\\small Departments of Mathematics and Statistics    \\\small Columbia University, New York, NY 10027
                            \\\small {\tt ik@math.columbia.edu}
\and CONSTANTINOS KARDARAS\\\small Department of   Statistics, Columbia University    \\\small  New York, NY 10027~                      ~ {\tt kardaras@stat.columbia.edu}
                             }

\begin{document}
\maketitle

\begin{abstract}
\noindent \small A financial market is called ``diverse"  if no
single stock is ever allowed to dominate the entire market in terms
of relative capitalization. In the context of the standard It\^
o-process model initiated by Samuelson (1965) we formulate this
property (and the allied, successively weaker notions of  ``weak
diversity" and ``asymptotic weak diversity") in precise terms. We
show that diversity is possible to achieve, but delicate. Several
illustrative examples  are provided,  which demonstrate  that
weakly-diverse financial markets contain relative arbitrage
opportunities: it is possible to {\it outperform} (or underperform)
such markets over
arbitrary time-horizons. The existence of such relative arbitrage
does not interfere with the development of option pricing, and has
interesting consequences for the pricing of long-term warrants and
for put-call parity. Several open questions are suggested for
further study.
\end{abstract}

\vspace{2mm}


\noindent
order statistics, local times, stochastic differential equations, strict local martingales.



\newpage
\input amssym.def
\input amssym


\section{Introduction}

\setcounter{equation}{0}
\setcounter{Assumption}{0}
\setcounter{Theorem}{0}
\setcounter{Proposition}{0}
\setcounter{Corollary}{0}
\setcounter{Lemma}{0}
\setcounter{Remark}{0}
\setcounter{Example}{0}

The notion of diversity for financial markets was introduced and
studied in the recent paper and monograph by Fernholz (1999, 2002).
It means, roughly, that  no individual stock is ever allowed to
dominate the entire market  in terms of relative capitalization. In
the context of the standard It\^ o-process,
geometric-Brownian-Motion-based  model for financial markets
introduced by Samuelson (1965), it is shown in the above-mentioned
monograph how to generate portfolios with ``good diversification"
properties in a systematic way and how to use these properties for
passive portfolio management. In particular, it is shown that
diversity in such models can lead to  arbitrage opportunities
relative to the market, over sufficiently long time-horizons.
The present paper complements this effort by showing that, under appropriate conditions, diversity in such models in indeed possible, but nonetheless rather delicate, to achieve. The conditions mandate, roughly, that the largest stock have ``strongly negative" rate of growth, resulting in a sufficiently strong repelling drift away from an appropriate boundary; and that all other stocks have ``sufficiently high" rates of growth.

\smallskip
Section 2 sets up the model and the notation that are used throughout the paper.  Section 3 introduces the ``market portfolio", in terms of which the notion of ``diversity" and the allied but successively weaker notions of ``weak diversity" and ``asymptotic weak diversity" are defined in section 4. The dynamics of the order statistics for the market-portfolio-weights (the so-called ``ranked market weights") are studied in section 5, and in terms of them sufficient conditions for diversity are established in section 6. These are illustrated by means of several examples, including models which  are weakly diverse but fail to be diverse.  Section 7 contains a model for which weak diversity fails on finite time-horizons, but prevails as the time-horizon becomes infinite in the ``asymptotically weak" sense of section 4.

\smallskip
We study in (4.4)-(4.5) and the Appendix a {\sl diversity-weighted}  portfolio,  that {\it outperforms} significantly any  weakly-diverse market    over sufficiently long time-horizons, thus leading to arbitrage relative to the market. In section 8 we introduce the  ``mirror portfolios" and study their properties; these are then used to show that in the context of a weakly-diverse market it is possible to  outperform (or underperform)  the market-portfolio over {\it arbitrary} time-horizons.

\smallskip
Finally, we show in section 9 that familiar techniques for option pricing  can be carried out in the context of markets that are diverse, despite the absence of an equivalent martingale measure. This has interesting ramifications for put-call parity, and for the prices of call-options over exceedingly long time-horizons: these are shown to approach zero, just as intuition would mandate, rather than the initial stock-price (as when an equivalent martingale measure exists for every finite time-horizon).  Several open problems are suggested.


\section{The Model}

    \setcounter{equation}{0}
     \setcounter{Assumption}{0}
    \setcounter{Theorem}{0}
     \setcounter{Proposition}{0}
     \setcounter{Corollary}{0}
     \setcounter{Lemma}{0}
    \setcounter{Remark}{0}

We shall place ourselves in the standard  It\^ o-process model for a financial market which goes back to Samuelson (1965). This model contains $n$ risky assets (stocks), with prices-per-share driven by $m$ independent Brownian motions  as follows:
\begin{equation}\label{2.1}
dX_i(t) = X_i(t) \left[ \, b_i(t)\,dt + \sum_{\nu=1}^m \sigma_{i \nu} (t)\,dW_{\nu} (t) \, \right] \, , ~~ i = 1, \ldots, n\,
\end{equation}
for $0 \le t < \infty$, with $m \ge n\,$. Here $X_i(t)$ stands for the price of the $i^{th}$ asset at time $t$, and  $W(\cdot) = \left(W_1(\cdot) , \ldots, W_m(\cdot) \right)'$ is a vector of $m$ independent  standard Brownian motions, the ``factors" of the model. These processes are defined on a  probability  space $(\Omega,\F,  P)$  and are adapted to a given filtration $ {\bf F} = \left\{ \F(t) \right\}_{0 \le t < \infty}\,$  with $\F (0) = \{ \emptyset, \Omega \}$ mod. $P$; this satisfies the ``usual conditions" (right-continuity, augmentation by   $P-$negligible sets), and may be strictly larger than the one generated by the driving   $m-$dimensional Brownian motion $\,W(\cdot)$.

\smallskip
The vector-valued process  $\,b(\cdot) = \bigl( b_1(\cdot), \ldots, b_n(\cdot) \bigr)'$ of {\it rates of return}, and the $(n \times m)-$matrix-valued process $\sigma(\cdot) = \bigl\{ \sigma_{i \nu}(\cdot) \bigr\}_{1 \le i \le n,\, 1 \le \nu \le m}\,$ of {\it volatilities}, are assumed to be  ${\bf F}-$progressively measurable and to satisfy almost surely (a.s.) the  conditions
\begin{equation}\label{2.2}
\sum_{i=1}^n
\int_0^T   (b_i (t) )^2   \,dt \, < \, \infty\,,~~~ \forall ~ ~T \in ( 0,\infty)\,,
\end{equation}
\begin{equation}\label{2.3}
\eps || \xi ||^2 \, \le \xi' \sigma (t) \sigma ' (t) \xi \, \le \,  M || \xi ||^2\, ,~~~ \forall~ ~t \in [0,\infty)~~~ \hbox{and}~~~ \xi \in \R^n\,,
\end{equation}
for some real constants $\,M > \eps >0$. We may re-write (\ref{2.1}) in the equivalent form
\begin{equation}\label{2.4}
d \Bigl( \log X_i(t) \Bigr) \,=\, \gamma_i(t) \,dt + \sum_{\nu = 1}^m \sigma_{i \nu} (t)\, dW_{\nu}(t)\,,~~~ i =1, \ldots, n\,.
\end{equation}
Here we have denoted by $\,
\gamma_i(t) := b_i(t) - \frac{1}{2} a_{ii}(t)\,$, $\, i =1, \ldots, n
\,$ the individual stock {\it growth-rates}, and by $a(\cdot) = \bigl\{ a_{ij}(\cdot) \bigr\}_{1 \le i, j \le n}\,$ the $(n \times n)-$matrix of variation-covariation rate processes
\begin{equation}\label{2.5}
a_{ij}(t)\,:=\, \sum_{\nu=1}^m \sigma_{i \nu} (t) \sigma_{j \nu} (t) \,=\, \Bigl( \sigma (t) \sigma' (t) \Bigr)_{ij}  \,=\,
\frac{d}{dt} \lan \log X_i, \log X_j \ran (t)\,.
\end{equation}

Placed in the above market-model $\,{\cal M}$ of (\ref{2.1})-(\ref{2.3}), an economic agent can decide what proportion $\pi_i(t)$ of his wealth to invest in each of the stocks $\,i=1, \ldots, n\,$ at every time $t \in [0,\infty)$. The resulting {\it portfolio process} $\, \pi(\cdot) = \bigl( \pi_1(\cdot), \ldots, \pi_n(\cdot) \bigr)'$ takes values in the set
\[
\Delta_+^n \, = \, \left\{ (\pi_1, \ldots, \pi_n) \in \R^n \, \Big| \, \pi_1 \ge 0, \ldots, \pi_n \ge 0 \quad \hbox{and} \quad \sum_{i=1}^n \pi_i = 1 \right\}\,,
\]
and is supposed to be $\, {\bf F}-$progressively measurable. Starting with initial capital $z>0$, the {\it value process} $\,Z^{\pi}(\cdot)$ of the portfolio $\pi(\cdot)$ satisfies the analogue
\begin{equation}
\label{2.6}
\frac{dZ^{\pi}(t)}{Z^{\pi}(t)} ~=~ \sum_{i=1}^n \pi_i(t) \cdot  \frac{dX_i(t)}{X_i(t)}  ~=~  b^{\pi }(t)\,dt + \sum_{\nu=1}^m \sigma_{\nu}^{\pi}(t) \, dW_{\nu}(t)\,, ~~ ~ Z^{\pi}(0) =z
\end{equation}
of the equation (\ref{2.1}), where
\begin{equation}
\label{2.7}
 b^{\pi }(t) :=  \sum_{i=1}^n \pi_i(t) b_i(t)\,,~~~ \sigma_{\nu}^{\pi}(t) := \sum_{i=1}^n \pi_i(t) \sigma_{i \nu}(t)
\end{equation}
for $\,\nu = 1, \ldots, m$, are respectively the rate-of-return and the volatility co\" efficients of the portfolio. By analogy with (\ref{2.4}), we may write the solution of the equation (\ref{2.6}) in the form
\[
d \Bigl( \log Z^{\pi}(t) \Bigr) \,=\, \gamma^{\pi}(t) \,dt + \sum_{\nu = 1}^m \sigma^{\pi}_{\nu} (t)\, dW_{\nu}(t)\,,~~~~~~~\hbox{with}
\]
\begin{equation}
\label{2.8}
\gamma^{\pi }(t) :=  \sum_{i=1}^n \pi_i(t) \gamma_i(t) + \gamma^{\pi }_*(t)\,, \quad
\gamma^{\pi }_*(t) :=  \frac{1}{2} \,               \left(
\sum_{i=1}^n \pi_i(t) a_{ii}(t) - \sum_{i=1}^n
\sum_{j=1}^n \pi_i(t) a_{ij}(t) \pi_j(t)                                 \right)~~~~
\end{equation}
denoting, respectively, the {\it growth-rate} and the {\it excess-growth-rate} of the portfolio $\pi(\cdot)$.

In order to set the stage for notions and developments that follow, let us introduce the ``order-statistics" notation for the weights
\begin{equation}
\label{3.2}
\max_{1\le i\le n} \pi_i(t) \, =: \, \pi_{(1)}(t) \, \ge \,
\pi_{(2)}(t)   \, \ge \,   \ldots   \, \ge \,     \pi_{(n-1)}(t)  \, \ge \,  \pi_{(n)}(t) \, =: \, \min_{1\le i\le n} \pi_i(t)
\end{equation}
 of  a portfolio $\, \pi(\cdot)$, ranked at time $t$ from the largest $\,\pi_{(1)}(t)$ to the smallest  $\,\pi_{(n)}(t)$.

We shall also introduce the following notion of {\bf relative arbitrage}: given any two portfolios $\, \pi(\cdot)$ and $\, \rho(\cdot)$ and a real constant $T>0$, we shall say that $\pi(\cdot)$ {\it represents an  arbitrage opportunity relative to $\rho(\cdot)$ over the time-horizon} $[0,T]$  if, starting with the same initial capital $\, Z^{\pi}(0) =  Z^{\rho} (0) = z >0\,$, we have
\begin{equation}
\label{AO}
P \left[ Z^{\pi}(T) \ge  Z^{\rho} (T) \right] \,=\,1\,~~~~\hbox{and}~~~~ ~
P \left[ Z^{\pi}(T) >  Z^{\rho}(T) \right] >0 \,.
\end{equation}


\section{The Market Portfolio}

\setcounter{equation}{0}
\setcounter{Assumption}{0}
\setcounter{Theorem}{0}
\setcounter{Proposition}{0}
\setcounter{Corollary}{0}
\setcounter{Lemma}{0}
\setcounter{Remark}{0}
\setcounter{Example}{0}

If we view the stock-price $X_i(t)$ as the capitalization of the $i^{th}$ company at time $t$, then the quantities
\begin{equation}
\label{3.1}
Z(t) := X_1(t) + \ldots + X_n(t) ~~~~~~ \hbox{and}~~~~~~ \mu_i(t) := \frac{X_i(t)}{Z(t)}\,,~~~~i =1, \ldots, n
\end{equation}
denote the total capitalization of the market  and the relative capitalizations of the individual companies, respectively. Since $~
0 < \mu_i(t) <1\,,~~ \forall ~ i=1, \ldots, n\,$ and $~\sum_{i=1}^n \mu_i(t) =1\,$, we may think of the vector process $\,\mu(\cdot) = \bigl( \mu_1(\cdot), \ldots, \mu_1(\cdot) \bigr)'$ as a portfolio rule that invests a proportion $\,\mu_i(t)$ of the current wealth $\,Z^{\mu}(t)$ in the $i^{th}$ asset, at all times $t \in [0, \infty)$. Then the resulting value-process $\,Z^{\mu}(\cdot)$ satisfies
\[
\frac{dZ^{\mu}(t)}{Z^{\mu}(t)} ~=~ \sum_{i=1}^n \mu_i(t) \cdot  \frac{dX_i(t)}{X_i(t)}  ~=~ \sum_{i=1}^n  \frac{dX_i(t)}{Z(t)}  ~=~ \frac{dZ(t)}{Z(t)}\,,
\]
as postulated by (\ref{2.6}) and (\ref{3.1}); and if we start with initial capital  $\,Z^{\mu}(0) = Z(0)$, we have $\,Z^{\mu}(\cdot) \equiv Z(\cdot)$, the total  market capitalization.  In other words, investing according to the portfolio  process $\,\mu(\cdot)$ amounts to ownership of the entire market. For this reason  we call $\,\mu(\cdot)$ the
{\it market portfolio} for $\,\M$.


\section{Notions of Diversity}

\setcounter{equation}{0}
\setcounter{Assumption}{0}
\setcounter{Theorem}{0}
\setcounter{Proposition}{0}
\setcounter{Corollary}{0}
\setcounter{Lemma}{0}
\setcounter{Remark}{0}
\setcounter{Example}{0}

The notion of ``diversity" for a financial market corresponds to the intuitive idea that no single company should be allowed to dominate the entire market in terms of relative capitalization. To make this notion precise, let us say that the model $\,\M$ of (\ref{2.1})-(\ref{2.3}) is {\tt diverse} on the time-horizon $\, [0,T]$, if there exists a number $\, \d \in (0,1)$ such that the quantities of (\ref{3.1}) satisfy almost surely
\begin{equation}\label{4.1}
\mu_{(1)}(t) < 1-\d\,, ~~~~ \forall ~~ 0 \le t \le T
\end{equation}

\noindent
in the notation of ({\ref{3.2}). In a similar vein, we say that $\,\M$  is {\tt weakly diverse} on the time-horizon $\, [0,T]$, if for some $\, \d \in (0,1)$ we have
\begin{equation}\label{4.2}
\frac{1}{T}\, \int_0^T
\mu_{(1)}(t) \, dt \,< \,  1-\d
\end{equation}
almost surely. We say that $\,\M$  is {\tt uniformly weakly diverse} over   $\, [T_0, \infty)$, if there exists a $\, \d \in (0,1)$ such that (\ref{4.2}) holds a.s. for every $\, T \in [T_0, \infty)$.
And  $\, \M$ is called {\tt asymptotically weakly diverse} if, for some $\, \d \in (0,1)$, we have almost surely:
\begin{equation}\label{4.3}
\overline{\lim}_{T \rightarrow \infty}\,
\frac{1}{T}\, \int_0^T
\mu_{(1)}(t) \, dt \,< \,  1-\d\,.
\end{equation}

These notions were introduced in the paper by Fernholz (1999) and are studied in detail in the recent monograph Fernholz (2002). In particular, it is shown in Example 3.3.3 of this book that {\it if the model} $\,\M$ {\it of (\ref{2.1})-(\ref{2.3}) is weakly diverse, then it contains arbitrage opportunities relative to the market portfolio.}

\medskip
We provide here another example of such an arbitrage opportunity,  in a weakly diverse market and for the so-called {\bf diversity-weighted  portfolio}  $\,\pi^{(p)}(\cdot) = \bigl( \pi_1^{(p)}
(\cdot), \ldots, \pi_n^{(p)}(\cdot) \bigr)'$. For some given $\, 0<p<1\,$, this is defined in terms of the market portfolio $\,\mu(\cdot)$ of (\ref{3.1}), by
\begin{equation}
\label{4.4}
\pi_i^{(p)} (t) \, := \, \frac{\bigl( \mu_i (t) \bigr)^p}{\sum_{j=1}^n
\bigl( \mu_j (t) \bigr)^p}\,\,,~~~~ \forall ~~ i=1, \ldots, n \,.
\end{equation}
Compared to $\,\mu(\cdot)$, the portfolio $\, \pi^{(p)}
(\cdot)$ in (\ref{4.4}) decreases the proportion(s) held in the largest stock(s) and increases those placed in the smallest stock(s), while preserving the relative rankings of all stocks. The actual performance of this portfolio relative to the S\&P500 index over a 22-year period is discussed in detail by Fernholz (2002), along with various issues of practical implementation.

We show in Appendix A that if the model $\,\mathcal{M}$ of (\ref{2.1})-(\ref{2.3}) is weakly diverse on a finite time-horizon $\, [0,T]$,
then, starting with initial capital  equal to $\, Z^{\mu}(0)$, the value-process $\,Z^{\pi^{(p)}}(\cdot)$ of the portfolio in (\ref{4.4}) satisfies
\begin{equation}
\label{4.5}
P \left[ Z^{\pi^{(p)}}(T) >  Z^{\mu}(T) \right] \,=\,1\,,~~~
\hbox{provided that} ~~~ T  \ge  \, \frac{2}{p \eps \delta} \cdot  \log n \,.
\end{equation}
In particular, $\, \pi^{(p)} (\cdot)$ is then an arbitrage opportunity relative to the market $\, \mu(\cdot)$, in the sense of (\ref{AO}).

\smallskip
{\it What conditions on the co\" efficients} $\,b(\cdot)$, $\sigma(\cdot)$ {\it of} $\,\M$ {\it are sufficient for guaranteeing diversity, as in (\ref{4.1})?} $~$ Certainly $\,\M$  cannot be diverse if $\,b_1(\cdot), \ldots, b_n(\cdot)$ are bounded uniformly in $(t, \omega)$, or even if they satisfy a condition of the Novikov type
\begin{equation}
\label{4.7}
E \left[ \exp \left\{  \frac{1}{2}\, \int_0^T  \Big| \Big| b(t) \Big| \Big|^2\,dt \right\} \right ] \,<\, \infty \,,~~~ \forall ~~ T \in (0,\infty)\,.
\end{equation}
The reason is that, under the condition (\ref{4.7}), the Girsanov theorem produces an equivalent probability measure $\,Q$  under which the price-processes $\,X_1(\cdot), \ldots, X_n(\cdot)$ in (\ref{2.1}) become martingales. This proscribes (\ref{AO}), let alone the equation of (\ref{4.5}), for \textit{any} $\,T \in (0, \infty)$; see Appendix A for an argument in a somewhat more general context.

On the other hand, if we forego the square-integrability condition (2.2), then there exist portfolios $\pi(\cdot)$ that lead to ``instantaneous arbitrage" $\, P \left[ Z^{\pi}(T) >  Z^{\mu}(T)\,,~ \forall \,\,T    \in (0,\infty) \right] \,=\,1\,$ relative to the market; see Appendix B.

\smallskip
We shall see in section 6 that  diversity is ensured by a strongly negative rate of growth for the largest stock, resulting in a sufficiently strong repelling drift (e.g., a log-pole-type singularity) away from an appropriate boundary, and by non-negative growth-rates for all the other stocks. It turns out, however, that diversity does not prohibit the familiar treatments of option pricing, hedging, or portfolio optimization problems in the context of diverse markets; we elaborate on this point in section 9.


\section{The Dynamics of Ranked Market-Weights}

\setcounter{equation}{0}
\setcounter{Assumption}{0}
\setcounter{Theorem}{0}
\setcounter{Proposition}{0}
\setcounter{Corollary}{0}
\setcounter{Lemma}{0}
\setcounter{Remark}{0}
\setcounter{Example}{0}

 A simple application of It\^ o's rule to the equation (\ref{2.4}) shows that we have
\begin{equation}
\label{5.2}
d \Bigl( \log \mu_i(t) \Bigr) ~=~ \left( \g_i(t) - \g^{\mu}(t) \right)  dt \, + \, \sum_{\nu=1}^m \left( \sigma_{i \nu} (t) -  \sigma_{ \nu}^{\mu} (t) \right)   dW_{\nu}(t)\,,~~~ i = 1, \ldots, n
\end{equation}
in the notation of (\ref{2.7}), (\ref{2.8}), or equivalently
\begin{equation}
\label{5.3}
\frac{d  \mu_i(t)}{\mu_i(t)} ~=~ \left( \g_i(t) - \g^{\mu}(t) + \frac{1}{2}\, \tau^{\mu}_{ii}(t) \right) \, dt \, + \, \sum_{\nu=1}^m \left( \sigma_{i \nu} (t) -  \sigma_{ \nu}^{\mu} (t) \right) \, dW_{\nu}(t)
\end{equation}
for $\, i = 1, \ldots, n\,$.  Here, by analogy with (\ref{2.5}), we have introduced
\begin{equation}
\label{5.4}
\tau^{\pi}_{ij}(t) \,:=\, \sum_{\nu=1}^m
\left( \sigma_{i \nu} (t) -  \sigma_{ \nu}^{\pi} (t) \right)
\left( \sigma_{j \nu} (t) -  \sigma_{ \nu}^{\pi} (t) \right)
\,=\, a_{ij}(t) - a^{\pi}_i (t) - a^{\pi}_j (t) + a^{\pi \pi} (t)\,,
\end{equation}
the relative covariance (matrix-valued) process of an arbitrary portfolio $\,\pi(\cdot)$, and set
\[
a^{\pi}_i (t) \,:=\, \sum_{k=1}^m \pi_{k}(t) a_{i k}(t)\,,~~~ ~~~
a^{\pi \pi} (t) \,:=\, \sum_{i=1}^n \sum_{k=1}^n\pi_i(t) a_{ik}(t) \pi_k(t)\,.
\]
In terms of the quantities of (\ref{5.4}) we can express the excess rate of growth of (\ref{2.8}) as
\begin{equation}
\label{6.7a}
\g^{\pi}_*(t) \,=\, \frac{1}{2} \sum_{i=1}^n \pi_{i}(t) \tau^{\pi}_{ii}(t)
\,;
\end{equation}
and for arbitrary portfolios $\,\pi(\cdot)$, $\,\rho(\cdot)$ we have the  ``num\' eraire-invariance" property
\begin{equation}
\label{NI}
\gamma^{\pi }_*(t)\, =\,  \frac{1}{2}  \left(
\sum_{i=1}^n \pi_i(t) \tau_{ii}^{\rho}(t) - \sum_{i=1}^n \sum_{j=1}^n \pi_i(t)  \pi_j(t) \tau^{\rho}_{ij}(t)\right)\,;
\end{equation}
see Lemmata 1.3.4 and 1.3.6 in Fernholz (2002).

Now
let us denote by $\,\{ p_t(1), \ldots, p_t(n)\}$ the random permutation of $\,\{1, \ldots, n\}$  for which
\begin{equation}
\label{OS}
\mu_{p_t(k)}(t) = \mu_{(k)}(t)\,,~~~~~~\hbox{and}~~~
p_t(k) < p_t(k+1)~~~ \hbox{if}~~ ~\mu_{(k)}(t) = \mu_{(k+1)}(t)\,,
\end{equation}
hold for $\,k =1, \ldots, n$. This means, roughly, that $\,p_t(k)$ is the name (i.e., index) of the stock with the  $\,k^{th}$ largest relative capitalization at time $\,t$, and that ``ties are resolved by resorting to the lowest index". Using It\^ o's rule for convex functions of semimartingales, it is shown in Fernholz (2001, 2002) that the ranked market-weights of (\ref{3.2}) satisfy the dynamics
\begin{eqnarray}
\label{5.6}
d \Bigl( \log \mu_{(k)}(t) \Bigr) ~&=&~
\left( \g_{p_t(k)}(t) - \g^{\mu}(t) \right) \, dt \, + \, \sum_{\nu=1}^m \left( \sigma_{p_t(k) \nu} (t) -  \sigma_{ \nu}^{\mu} (t) \right) \, dW_{\nu}(t) ~~~~~~~      \\
& & ~~~~~~~~~~~ + \, \frac{1}{2} \cdot \left[
d \Lambda^{(k, k+1)}  (t)
- d \Lambda^{ (k-1, k)} (t) \right]\,. \nonumber
\end{eqnarray}

\medskip
\noindent Here, for each $\,k=1, \ldots, n-1$, the quantity $\, \Lambda^{(k, k+1)}  (t)  :=
\Lambda_{\log \mu_{(k)} - \log \mu_{(k+1)}}(t) $  is the local time that the non-negative semimartingale $\, \log \left( \mu_{(k)} / \mu_{(k+1)} \right)(\cdot)$ has  accumulated at the origin by calendar time $\,t\,$; and we set $\,
\Lambda_{\log \mu_{(0)} - \log \mu_{(1)}}(\cdot) \equiv 0\,$ and $~ \Lambda_{\log \mu_{(n)} - \log \mu_{(n+1)}}(\cdot) \equiv 0\,$.

\smallskip
On the event $\,\{ \mu_{(1)}(t) > 1/2  \}$      we have   $\, \mu_{(2)}(t) < 1/2 \,$,  thus $\,
\int_0^{\infty} 1_{ \{ \mu_{(1)}(t) > 1/2  \}}\,
d \Lambda^{(1,2)}  (t) =0\,$. Therefore, with $\,k=1$ the equation (\ref{5.6}) reads
\begin{equation}
\label{5.7}
d \Bigl( \log \mu_{(1)}(t) \Bigr) \,=\,
\left( \g_{(1)}(t) - \g^{\mu}(t) \right) \, dt \,
+ \, \frac{1}{2} \cdot 1_{\left\{ \mu_{(1)}(t) \le 1/2 \right\}} \cdot
  d \Lambda^{(1,2)}  (t)\, + \,
\sqrt{\tau^{\mu}_{(11)}(t)} \cdot dB(t) ~~~~
 \end{equation}
where $\,B(\cdot)$ is standard Brownian motion and
\begin{equation}\label{5.8}
\g_{(k)} (t)\,:=\, \g_{p_t(k)}(t)\,,~~~~ \tau^{\mu}_{(kk)} (t) \,:=\,
\tau^{\mu}_{ii} (t) \Big|_{i=p_t(k)} ~.
\end{equation}

\medskip
\noindent  {\sl 5.1 Remark:}  $\,$ For a portfolio $\, \pi(\cdot)\,$ the conditions of (\ref{2.3}) lead to the inequalities
\begin{equation}
\label{ULB}
 \varepsilon \Bigl( 1 - \pi_i (t) \Bigr)^2
\, \le \,
\tau^{\pi}_{ii} (t)
\, \le  \,
M\, (1-\pi_i (t))\, (2-\pi_i (t))
\end{equation}
for the quantities of (\ref{5.4}), and in the case of the market-portfolio to
\begin{equation}
\label{5.9}
\eps \, \Bigl( 1 - \mu_{(1)}(t) \Bigr)^2 \, \le \, \, \tau^{\mu}_{(kk)} (t) \, \le \, 2M\, ,~~~~~ t \ge 0\,,~ ~ k =1, \ldots, n\,.
\end{equation}
On the other hand, we show in  Appendix A that the inequalities of (\ref{2.3}) imply the bounds
\begin{equation}
\label{5.10}
\frac{\eps}{2}  \, \Bigl( 1 - \pi_{(1)}(t) \Bigr)  \, \le \, \, \gamma^{\pi}_{*} (t) \, \le \, M\,\Bigl( 1 - \pi_{(1)}(t) \Bigr)\, ,~~~~~0 \le  t < \infty
\end{equation}
in the notation of (\ref{2.8}), (\ref{3.2}).


\section{Ensuring Diversity}

\setcounter{equation}{0}
\setcounter{Assumption}{0}
\setcounter{Theorem}{0}
\setcounter{Proposition}{0}
\setcounter{Corollary}{0}
\setcounter{Lemma}{0}
\setcounter{Remark}{0}
\setcounter{Example}{0}

Suppose that we select a number $\, \d \in \Bigl( 0, 1- \mu_{(1)}(0)   \Bigr)$, where $\, \mu_{(1)}(0) = \max_{1 \le i \le n} X_{i}(0)/(X_1(0)+\cdots+X_n(0))$, and ask under what conditions we might have
\begin{equation}\label{6.1}
\mu_{(1)}(t)\,<\, 1- \d\,,~~~~ ~ \forall ~~0 \le t < \infty
\end{equation}
almost surely; this condition implies the requirement (\ref{4.1}) of diversity on any finite time-horizon $\,[0,T]$. To simplify the analysis  we shall assume $ \,\frac{1}{2} \le
\mu_{(1)}(0) < 1-\d\,$  and consider \begin{equation}
\label{6.2}
R := \inf \left\{ t \ge 0\,\Big|\, \mu_{(1)}(t) \le \frac{1}{2} \right\}\,,~~~~~ S := \inf \left\{ t \ge 0\,|\, \mu_{(1)}(t) \ge 1- \d \right\}\,,
\end{equation}
as well as the stopping times
\begin{equation}
\label{6.3}
 S_k := \inf \left\{ t \ge 0\,|\, \mu_{(1)}(t) \ge 1- \d_k \right\}\,,~~~~~\d_k = \d + \frac{1}{k}\,,
\end{equation}
for all $\, k \in {\bf N}$ sufficiently large. For diversity, it will be enough to guarantee
\begin{equation}\label{6.4}
\overline{\lim}_{k \rightarrow \infty} P\left[ S_k < R \right] \,=\,0\,\,;
\end{equation}
because then $\,P[S<R] \, \le \, \overline{\lim}_{k \rightarrow \infty} P\left[ S_k < R \right] \,=\,0\,$, and this leads to (\ref{6.1}).

\bigskip
\noindent
{\bf 6.1 THEOREM:}  {\it Suppose that on the event} $\, \left\{ \frac{1}{2} \le \mu_{(1)}(t) < 1-\d \right\}$ {\it we have}
\begin{equation}
\label{6.5}
\g_{(k)}(t) \, \ge \, 0 \, \ge \, \g_{(1)}(t)\,,~~~~ \forall ~~ k =2, \ldots, n
\end{equation}
\begin{equation}
\label{6.6}
\min_{2 \le k \le n}\g_{(k)}(t) - \g_{(1)}(t) + \frac{\eps  }{2}~\, \ge \,
~\frac{M }{\d Q(t)}\,
\,,~~~~~\hbox{where}~~~~~~Q(t):= \log\left( \frac{1-\d}{\mu_{(1)}(t)}
\right)~ \,.
\end{equation}
{\it Then (\ref{6.4}), (\ref{6.1}) are satisfied. On any given, finite time-horizon $\,[0,T]$ the market is diverse and   $\, \int_0^T  Q^{-2} (t)\,dt < \infty\,$ holds a.s.}

\medskip
\noindent {\sl 6.1 Remark:} The condition (\ref{6.6}) holds, in particular,  if all stocks but the largest have non-negative growth rates, whereas the growth rate of the largest stock is negative and exhibits a log-pole-type singularity as the relative capitalization of the largest stock approaches $\,1-\d\,$: $\,
\g_{(1)}(t) \, \le \, - \frac{M }{\d Q(t)} \,$ on the event $\,  \{ 1/2 \le \mu_{(1)}(t) < 1-\d  \}\,$.

\medskip
\noindent
{\sl 6.2 REMARK:} $\,$
In terms of our market-model $\, \M$ of Section 2  we may specify, for instance, a non-random volatility matrix $\,\sigma = \left\{ \sigma_{ i \nu } \right\}_{1\le i \le n,\, 1 \le \nu \le m}\,$ with the properties (\ref{2.3}) as well as a vector $\, g =(g_1, \ldots, g_n)'$ of non-negative numbers, and impose (\ref{2.4}) in the form of a system of stochastic differential equations
\begin{equation}
\label{2.4a}
d \Bigl( \log X_i(t) \Bigr)
\,= \, \left\{
g_i \cdot 1_{{\cal O} _i^c}(X(t))
- \frac{M}{\d} \cdot \frac{1_{{\cal O} _i}(X(t)) }{\log\left( \frac{1-\d}
{X_{i}(t)} \sum_{j=1}^{n} X_j(t) \right)}
\right\}\, dt\,
+\, \sum_{\nu = 1}^m \sigma_{i \nu}  dW_{\nu}(t)~~~~
\end{equation}
for the vector  of stock-price processes $X(\cdot) = \bigl( \,X_1 (\cdot), \ldots ,X_n (\cdot) \bigr)'$ . We are using here the notation
\[
{\cal O}_1 := \left\{ x \in (0,\infty)^n\,|\,\, x_1 \ge \max_{2 \le j \le n}x_j \right\}\,,~~~ ~ {\cal O}_n := \left\{ x \in (0,\infty)^n\,|\, \,x_n > \max_{1 \le j \le n-1}x_j \right\}\,,
\]
\[
{\cal O}_i := \left\{ x \in (0,\infty)^n\,|\, \,x_i > \max_{1 \le j \le i-1}x_j\,,~ x_i \ge \max_{i+1 \le j \le n}x_j \right\}\,,~~~
\hbox{for}~~i=2, \ldots, n-1
\]
in order to keep track of the name of the stock with the largest capitalization in accordance with the convention of (\ref{OS}): $\, X(t) \in {\cal O}_i\, \Leftrightarrow \,p_t(1) =i \,$.  With this specification all stocks but the largest behave like geometric Brownian motions  (with growth rates $\, g_i \ge 0\,$ as long as $\,i \ne p_t (1)\,$, and variances $\,\sum_{\nu =1}^{m} \sigma_{i \nu }^{2}\,$), whereas the log-price of the largest stock is subjected to a log-pole-type singularity in its drift, away from an appropriate right-boundary. Standard theory (see Veretennikov (1981)) guarantees that the system of (\ref{2.4a}) has a pathwise unique, strong solution $\,X(\cdot)$  on each interval of the form $ {\bf [}$$0,S_k$${\bf ]}$, for all $\,k \in {\bf N}$ sufficiently large, and thus also on ${\bf [}$$0,S$${\bf )}$ = ${\bf [}$$0, \infty $${\bf )}\, $ by the Theorem. The equations (\ref{2.4a}) prescribe rates of return
\[
b_i (t)\,=\, \frac{1}{2} a_{ii} + g_i \cdot 1_{{\cal O} _i^c}(X(t))
- \frac{M}{\d} \cdot \frac{1_{{\cal O} _i}(X(t)) }{\log\left( \frac{1-\d}
{X_{i}(t)} \sum_{j=1}^{n} X_j(t) \right)}\,, ~~~~~\, i = 1, \cdots, n\,
\]
for the model of (\ref{2.1}), (\ref{2.5}). From the last assertion of Theorem 6.1 these rates satisfy $\, \sum_{i=1}^n \int_0^T (b_i (t) )^2\, dt < \infty\,$ a.s., which is the requirement (\ref{2.2}).

\medskip
\noindent {\sl Proof of Theorem 6.1:} On the event $\, \left\{ \frac{1}{2} \le \mu_{(1)}(t) < 1-\d \right\}$ under consideration  the conditions of (\ref{6.5}) and (\ref{6.6}) lead to
\begin{eqnarray}
\label{6.7}
\g^{\mu}(t) -\g_{(1)}(t)
\,&=&\,
\sum_{k=1}^n \mu_{(k)}(t) \g_{(k)}(t) - \g_{(1)}(t) + \g^{\mu}_*(t)
\\
\nonumber
&=& \sum_{k=2}^n \mu_{(k)}(t) \g_{(k)}(t)  - \left(1- \mu_{(1)}(t)\right) \g_{(1)}(t)   + \frac{1}{2} \sum_{k=1}^n \mu_{(k)}(t) \,\tau^{\mu}_{(kk)} (t)
\\
\nonumber
& \ge & \left(1- \mu_{(1)}(t)\right) \left( \min_{2\le k \le n} \g_{(k)}(t) -\g_{(1)}(t) \right) + \frac{\eps}{2} \cdot \left(1- \mu_{(1)}(t)\right)
 \\
\nonumber
& \ge & \d \, \left[  \, \min_{2\le k \le n} \g_{(k)}(t) -\g_{(1)}(t) + \frac{\eps  }{2} \, \right] \, \ge \,
 \frac{M}{Q(t)}
\,\,,
\end{eqnarray}
almost surely, with the help of (\ref{6.7a}), (\ref{5.10}) and (\ref{6.1}).  For the process $\, Q(\cdot)$ of (\ref{6.6}) we have from It\^ o's rule and (\ref{5.7}) the semimartingale decomposition
\begin{equation}
\label{6.8}
d ( \log Q(t))    =  \frac{1}{Q(t)}
\left( \g^{\mu}(t) -\g_{(1)}(t) -
\frac{\tau^{\mu}_{(11)}(t)}{2Q(t)} \right) dt
- \frac{\sqrt{\tau^{\mu}_{(11)}(t)}}{Q(t)}\, dB(t)
 +
\frac{1_{ \{ \mu_{(1)} (t) \le 1/2 \} }}{Q(t)} \, d \Lambda^{(1,2)} (t)\,;
 \end{equation}
  in conjunction with  (\ref{6.7}) and the second inequality in (\ref{5.9}), this gives
\[
  \log \frac{Q(\ell \wedge R \wedge  S_k)}{ Q(0)  }
 ~\ge~   \int_0^{\ell \wedge R \wedge S_k} \left( \frac{ 2M -
\tau^{\mu}_{(11)}(t)}{2Q^2(t)}\right)\,dt \,-\, \int_0^{\ell \wedge T \wedge S_k} \frac{1}{Q(t)}\,
 \sqrt{\tau^{\mu}_{(11)}(t)}\cdot dB(t)
\]
\begin{equation}
\label{6.9}
\ge \,~   - \int_0^{\ell \wedge R \wedge S_k}  \frac{1}{Q(t)}\,
 \sqrt{\tau^{\mu}_{(11)}(t)}\cdot dB(t)
\end{equation}
almost surely, for all integers $\,\ell$ and $\,k$ large enough.

\smallskip
Now let us take expectations in (\ref{6.9}). On the event $\,\{ t \le R \wedge S_k \}\,$ we have
\[
\eps \d \, \le \, \tau^{\mu}_{(11)}(t) \, \le \, 2M \,,~~~~
\log\left(  \frac{1-\d}{1-\d_k} \right)\,\le\, Q(t) \,\le\, \log\left(  \frac{1-\d}{1/2} \right)\,
\]
from  (\ref{5.10}) and (\ref{6.1})-(\ref{6.3}), (6.6). These bounds imply that the expectation of the stochastic integral is equal to zero. We are led to the inequalities
\begin{eqnarray*}
\log (Q(0)) \,&\le &\, E \left[\, \log \left( Q \left( \ell \wedge R \wedge S_k \right) \, \right)\right]       \\
& \le & \, \log \log \left( \frac{1-\d}{1-\d_k} \right) \cdot
P \left[ S_k < \ell \wedge R \right] \,+\,
\log \log \left( \frac{1-\d}{1/2} \right) \cdot
P \left[  \ell \wedge R  \le S_k \right] \,,
\end{eqnarray*}
and letting $\, \ell \rightarrow \infty$ we obtain
\begin{equation}\label{6.10}
- \log \log \left( \frac{1-\d}{1-\d_k} \right) \cdot
P \left[ S_k <  R \right] \,\le\,  - \log \log \left( \frac{1-\d}{\mu_{(1)}(0)} \right)\,+\,
\log \log \left( 2(1-\d) \right) \cdot
P \left[   R  \le S_k \right] \,.
\end{equation}
This inequality is valid for all $\,k \in {\bf N}$ sufficiently large. Finally, we divide by the number $\,\,- \log \log \left( \frac{1-\d}{1-\d_k} \right) >0\,$ in (\ref{6.10}), and then let $\,k \rightarrow \infty$; the desired conclusion (\ref{6.4}) follows.

\medskip
Now from (\ref{6.8}) the quadratic variation of the semimartingale $\, \log Q(\cdot)\,$ satisfies
\[
\eps \delta^2 \int_0^T \frac{1}{Q^2 (t)} \, dt \,<\,
\int_0^T \frac{\tau^{\mu}_{(11)}(t)}{Q^2 (t)} \, dt \,=\,
\langle \log Q \rangle (T)  \,<\, \infty\,,~~~~\hbox{a.s.}
\]
in conjunction with (\ref{5.9}) and (\ref{6.1}), and the last claim of the theorem follows.~~~~~~~~~~~~~~~~~~~$\diamondsuit$

\bigskip
\smallskip
The part of this proof leading up to (\ref{6.10}) is similar to the argument used to establish the non-attainability of the origin by Brownian motion in dimension $\,n \ge 2$; see, for instance, pp.161-162 in Karatzas \& Shreve (1991)  and Stummer (1993). The fact that a pole-type singularity creates opportunities for relative arbitrage is reminiscent of an example due to A.V. Skorohod (e.g. Karatzas \& Shreve (1998), p.11), or of the work by Delbaen \& Schachermayer (1995) and by Levental \& Skorohod (1995).

\bigskip
\noindent {\sl 6.3 REMARK:} $\,$ The inequality of condition (\ref{6.6}) can be replaced by
\begin{equation}\label{6.6a}
\min_{2 \le k \le n}\g_{(k)}(t) - \g_{(1)}(t) + \frac{\eps  }{2}~\, \ge \, ~\frac{M }{\d} \cdot F ( Q(t)) \,,
\end{equation}
where $\, F: (0,\infty) \rightarrow (0, \infty)$ is a continuous function  with the  property that the associated  scale function
\begin{equation}
\label{6.11}
U(x) \,:=\, \int_1^x \exp \left[   - \int_1^y
F(z) dz \right] \, dy \,,~~~~~  0 < x < \infty
\end{equation}
satisfies $\,
U(0+) \,=\, - \infty\,$.
For instance,   $\,U(x) = \log x\,$  when $\,F(x)=1/x$ as in (\ref{6.6}) or (\ref{6.7}).

\medskip
The function $\,U(\cdot)$ of (\ref{6.11}) is of class $\, {\cal C}^2 \bigl( 0, \infty \bigr)$, so we can apply It\^ o's rule to the process $\, U(Q(t))$, $0 \le t < S$ as in (\ref{6.8}). Using the strict increase and strict concavity properties $\,U'(\cdot) >0$, $\,U''(\cdot) <0$ of the scale function in (\ref{6.11}), as well as the equation $\, U''(\cdot) + F(\cdot) U'(\cdot) = 0$, we can now repeat the steps of the argument that leads to the analogue
\begin{displaymath}
- U \left(  \log  \frac{1-\d}{1-\d_k} \right) \cdot
P \left[ S_k <   R \right] \,\le\,  - U \left( \log
 \frac{1-\d}{\mu_{(1)}(0)} \right)\,+\, U \Bigl(
\log ( 2(1-\d) \Bigr) \cdot
P \left[   R  \le S_k \right]
\end{displaymath}
of (\ref{6.10}), and hence to (\ref{6.4}) with the help of the requirement 
$\, U(0+) \,=\, - \infty\,$.


\section{An Asymptotically Weakly Diverse Market}

\setcounter{equation}{0}
\setcounter{Assumption}{0}
\setcounter{Theorem}{0}
\setcounter{Proposition}{0}
\setcounter{Corollary}{0}
\setcounter{Lemma}{0}
\setcounter{Remark}{0}
\setcounter{Example}{0}

Suppose we have a two-stock market model of the form
\begin{equation}
\label{D.1}
dX_i (t)\, =\, X_i(t)\, \left[ \, b_i (t)\, dt\,+\, \frac{1}{\sqrt{2}} \, dW_i(t)\, \right] \,,~~~ X_i (0) = x \, \in \, (0,\infty)~~~~~~\hbox{for}~~i=1,2\,
\end{equation}
driven by the planar Brownian Motion $\, W=(W_1, W_2)\,$. Then $\, W\,:=\, \frac{1}{\sqrt{2}} \, \bigl( W_2 - W_1 \bigr)\,$ is standard Brownian motion, and we have
\begin{equation}
\label{D.2}
X_2 (t) \,=\, X_1(t) \cdot \exp \bigl( Z(t) \bigr)\,,~~~~~\hbox{where}~~~~ Z(t) \,:=\, \int_0^t \bigl( b_2 (s) - b_1 (s) \bigr)\, ds \,+\, W(t)\,,
\end{equation}
\begin{equation}
\label{D.3}
\mu_1 (t) \,=\, \frac{X_1 (t)}{X_1 (t) +X_2 (t)}
\,=\, \frac{1}{1 +e^{\,Z(t)}}\,,~~ \mu_2 (t)
\,=\, \frac{1}{1 +e^{\,-Z(t)}}\,,~~~ \hbox{thus}~~~
\mu_{(1)} (t) \,=\, \frac{1}{1 +e^{\,-|Z(t)|}}
\end{equation}
for $\, 0 \le t < \infty\,$. Now let us select $\, b_1 (\cdot) \equiv 0\,$ and $\, b_2 (\cdot) \equiv - \alpha \, Z(\cdot)\, 1_{[1, \infty)} (\cdot)\,$ for a suitable real constant $\, \alpha >0\,$ to be determined below. With these choices the process $\, Z(\cdot)\,$ of (\ref{D.2}) becomes $\, Z(t)\,=\, W(t)\,$ for $\, 0 \le t \le 1\,$ and
\begin{equation} \label{D.4}
Z(t) \,=\, W(1)\,-\, \alpha \, \int_1^t Z(s)\,ds\,+\, \widetilde{W} (t)~~~~~\hbox{for  }~ 1 \le t < \infty\,,
\end{equation}
where $\, \{ \widetilde{W} ( t)\,:=\, W  ( t) - W(1)\,,~ 1 \le t < \infty\, \}$ is standard Brownian Motion and independent of $\, Z(1)\,=\, W(1)\,$. In other words, the process $\, \{ Z ( t)\,,~1 \le t < \infty\, \}$ is  Ornstein-Uhlenbeck, with gaussian initial distribution $\, \mathcal{N}$$ (0,1)\,$ and gaussian invariant distribution $\, \mathcal{N}$$(0,1/2 \alpha)\,$; see  Karatzas \& Shreve (1991), page 358 for the latter assertion. With the choice $\, \alpha = 1/2\,$ the process $\,Z(\cdot)\,$ is {\it stationary}, and its ergodic behavior gives
\[
\lim_{T \rightarrow \infty} \frac{1}{T}\, \int_1^{T+1} \mu_{(1)}(t)\, dt
\, = \,
\lim_{T \rightarrow \infty} \frac{1}{T}\, \int_0^T
\frac{dt}{1 +e^{\,-|Z(t+1)|}}
\,=\,
E \left( \frac{1}{1 +e^{\,-|Z( 1)|}} \right)\,<\, 1 - \delta\,,~~~~~~~ \hbox{a.s.}
\]
for any $\, 0 < \delta < E \left( \frac{e^{\,-|Z( 1)|}}{1 +e^{\,-|Z( 1)|}} \right) \,=\,
\sqrt{\frac{2}{\pi}}\, \int_0^{\infty}
\frac{e^{\,-z}}{1 +e^{\, -z}} \,e^{\,-z^2 /2}\, dz\,$. Thus, {\sl the model $\, \mathcal{M}\,$ of (\ref{D.1}) is  asymptotically weakly diverse.}

However, {\it diversity  fails for this model}. For any $\,T \in [1, \infty)\,$ and $\, \delta \in (0,\infty)\,$ we have
\[
P \left[ \, \mu_{(1)} (T) \ge 1 - \delta \, \right] \,=\,
P \left[ \, |Z(T)| \ge \xi \, \right] \,=\,
\frac{2}{\sqrt{2 \pi}} \int^{\infty}_{\xi} e^{\,-u^2/2}\,du
 \,>\,0\,,
\quad
\xi := \log \left( \frac{1 - \delta}{\delta} \right)\,.
\]
In fact, {\it weak diversity fails as well}. For an arbitrary $\,T \in (1, \infty)\,$ and $\, \delta \in (0,1)\,$, select $\, \eps \in (0,T)\,$ and $\, \zeta  >0\,$ so that $\,
\delta \, \ge \, \frac{( \eps / T ) + e^{\,- \zeta  }}{1 + e^{\,-\zeta  }}\,$; then it is straightforward that the event $\, A_{\eps, \zeta  }\,:=\, \left\{ \inf_{\eps \le t \le T} |Z(t)| \,\ge\, \zeta   \right\}\,$ has positive probability $\,P \left( A_{\eps, \zeta   } \right) >0\,$ and that
\[
\frac{1}{T}\, \int_{\eps}^T \mu_{(1)}(t)\, dt
\, = \,
\frac{1}{T}\, \int_{\eps}^T
\frac{dt}{1 +e^{\,-|Z(t)|}}
\, \ge \,
\frac{T-\eps}{T \,(1+e^{\,- \zeta   })}
\, \ge \, 1 - \delta ~~~~~~~~\hbox{holds a.e. on}~~ A_{\eps, \zeta   }\,
\]
thus leading to $\, P \left(  \, \int_{0}^T \mu_{(1)}(t)\, dt
\, \ge \, (1 - \delta) T \,  \right) \,>\,0\,$. It can be shown that the model of (\ref{D.1}) admits a unique equivalent martingale measure.

\medskip
\noindent {\sl  7.1 REMARK:} $\,$ The examples of section 6 can be easily modified  to produce {\sl a model $\, \mathcal{M}\,$ which is weakly diverse but not diverse.} Indeed, let us start by considering a model $\, \mathcal{M}^{({\rm 2} \delta)}\,$ with constant volatilities $\, \sigma_{ij}\,$ and with rates of return $\,b_i^{(2 \delta)}(\cdot)\,,~ i=1, \cdots, n\,$ such that $\,
P \bigl(\, \mu_{(1)} (t) < 1 - 2 \delta \,, ~\forall\,  0 \le t \le T \, \bigr) \, =\,1\,$
is satisfied for some $\,T \in (0,\infty)\,$ and $\, \delta \in (0, 1/4)\,$. The idea is to divide the time-horizon $\,[0,T]\,$ into the two intervals $\,[0,T/2)\,$ and $\,[T/2,T]\,$, select $\, \eta \in (2 \delta, 1/2)\,$, and set
\begin{equation} \label{D.5}
b_i (t) \, :=\, b_i^{(2 \delta)} (t) \cdot 1_{\{ S \le t \le T\,,~ S \le T/2 \}}
\,,~~~~~\hbox{where}~~~~~~~
S \, :=\, \inf \{ \, t \ge 0 \,|\, \mu_{(1)} (t) \ge 1 - \eta \, \} \,
\wedge \,T\,.~~~~
\end{equation}
We claim that the model $\, \mathcal{M}\,$, with  volatilities $\, \sigma_{ij}\,$ and rates of return given by (\ref{D.5}), is weakly diverse on $\,[0,T]\,$. To see this, consider two cases: For $\, \omega \in \{ S \le T/2 \}\,$  the recipe  (\ref{D.5}) and (\ref{4.1}) guarantee $\, \mu_{(1)} (t, \omega) < 1 - 2 \delta < 1 -  \delta \,$,  $\, \forall \, 0 \le t \le T\,$; and for $\, \omega \in \{ S > T/2 \}\,$ we have
\[
\frac{1}{T} \int_0^T  \mu_{(1)} (t, \omega) \,dt \, \le \,
\frac{1}{T} \int_0^{T/2} ( 1 - \eta) \,dt \, + \,
\frac{1}{T} \int_{T/2}^T 1 \cdot dt \, = \, 1 - ( \eta / 2) \,<\,1 - \delta\,.
\]

\smallskip
But for this $\, \mathcal{M}\,$  {\sl the property (\ref{4.1}) fails:} the event $\, B\,:=\, \{  S > T/2 \}\,$ has positive probability, and with $\, A \,:=\, \left\{ \, \max_{0 \le t \le T} \mu_{(1)} (t)\,\ge\, 1 - \delta\,\right\}\,$ we have  $\, P (A \cap B)>0\,$. To see this, consider the special case $\,n=2\,$, $\, \sigma_{12} = \sigma_{21} =0\,$, $\, \sigma_{11} =\sigma_{22} = 1 / \sqrt{2}\,$  as in (\ref{D.1}), and observe that on the event $\, B\,=\, \{  S > T/2 \}\,$ we have $\, Z(\cdot) \equiv W(\cdot)\,$ in (\ref{D.2}) and
\[
\max_{0 \le t \le T} \mu_{(1)} (t)\,\ge\, 1 - \delta~~~~~~~ \Longleftrightarrow~~~~~~~
\max_{0 \le t \le T} |Z (t)| \, \ge \, K \,:=\, \log \left( \frac{1-\delta}{\delta} \right) \,.
\]
Consequently,
\[
 P(A \cap B) \,=\, P \Bigl[ \,
\max_{0 \le t \le T} |W (t)| \, \ge \, K \,;\, S > T/2 \,\Bigr]
\, \ge \, P \Bigl[ \,
\max_{T/2 \le t \le T} |W (t) - W(T/2)| \, \ge \, 2K \,;\, S > T/2 \,\Bigr]
\]
\[
= \,
P \left(  \max_{T/2 \le t \le T} |W (t) - W(T/2)| \, \ge \, 2K \, \Big \vert \,  S > T/2 \right) \cdot P ( S > T/2 )~~~~~~~~~
\]
\[
\ge \,
P \left( \,  \max_{0 \le t \le T/2 } |W (t)| \, \ge \, 2K \,   \right) \cdot P ( S > T/2 ) \,>\, 0\,,~~~~~~~~~~~~~~~~~~~~~~~~~~~
\]

\smallskip
\noindent
since $\, \{ W (t) - W(T/2);\, T/2 \le t < \infty\,\}\,$ is a Brownian Motion and independent of $\, \F (T/2)\,$, a $\sigma-$algebra that  contains the event $\,\{ S > T/2\}$.


\section{Mirror Portfolios, Short-Horizon Relative Arbitrage}

\setcounter{equation}{0}
\setcounter{Assumption}{0}
\setcounter{Theorem}{0}
\setcounter{Proposition}{0}
\setcounter{Corollary}{0}
\setcounter{Lemma}{0}
\setcounter{Remark}{0}
\setcounter{Example}{0}

We saw in (\ref{4.5}) that in weakly diverse markets and over sufficiently long time-horizons, there exist portfolios (e.g., the diversity-weighted portfolio  $\, \pi^{(p)}(\cdot)$ of (\ref{4.4})) that represent  arbitrage opportunities relative to the market portfolio $\mu(\cdot)$. It is an open question whether this can be done over arbitrary, possibly small, time-horizons. We shall show in this section that, on arbitrary time-horizons, relative arbitrage can be constructed in the reverse direction: if short-selling is allowed, there always exist portfolios that constistently underperform a weakly diverse market, i.e., with respect to which the market-portfolio represents an arbitrage opportunity.

In order to do this we have to introduce the notion of {\bf extended portfolio}: a progressively measurable and uniformly bounded process $\, \pi(\cdot) = ( \pi_1(\cdot) , \cdots, \pi_n (\cdot) )'\,$ with values in $\, \Delta^n = \{ ( \pi_1 , \cdots, \pi_n )  \in \R^n\,|\, \sum_{i=1}^n \pi_i = 1 \}\,$. In other words, an extended portfolio can sell one or more stocks short, but certainly not all. By contrast, the portfolios of section 2 are ``all-long" portfolios: they allow no short-selling.

Let us fix a baseline portfolio $m(\cdot)$; this will typically, though not necessarily, be the market portfolio $\,\mu(\cdot)$. For any extended portfolio $\pi(\cdot)$ and any fixed real number $p \neq 0$ we define the $ \,p-${\sl mirror-image of $\pi(\cdot)$ with respect to} $\mu(\cdot)$ by
\begin{equation}
\label{B.1}
\widetilde \pi^{(p)} (\cdot)  \,:=\, p \, \pi (\cdot) + (1-p) \, m (\cdot) \,.
\end{equation}
This is clearly an extended portfolio, and  a portfolio in the strict sense of section 2
if this is the case  for $\pi(\cdot)$ and $0 < p < 1$. If $p=-1$ we call $\, \pi^{(-1)} (\cdot)  = 2 m (\cdot)- \pi (\cdot) \,$ the ``mirror image" of $\pi(\cdot)$ with respect to $m(\cdot)$. We notice
\begin{equation}
\label{B.2}
\left( \widetilde \pi^{(p)} \right)^{ \widetilde{\,}\, (q)} \,=\, \widetilde \pi^{(pq)}\,\,,~~~~~~~
\left(  \widetilde \pi^{(p)} \right)^{\widetilde{\,} \,(1/p)} \,=\,   \pi \,.
\end{equation}
Let us recall the notation $\, \tau^m (\cdot) = \left\{  \tau^m_{ij}  (\cdot) \right\}_{1 \le i,j \le n}\,$ of (\ref{5.3}) for the matrix-valued covariance process of $\, m (\cdot)$,   define the {\it relative covariance
of $\pi(\cdot)$ with respect to} $m (\cdot)$ by
\begin{equation}
\label{B.3}
\tau^m_{\pi \pi}(t)\,:=\, ( \pi (t) - m(t) )'\, a(t)\, ( \pi (t) - m(t) )\,\ge\, \varepsilon \, || \pi(t) - m(t) ||^2\,\,,
\end{equation}
and make the elementary observations
\begin{equation}
\label{B.4}
\tau^m (\cdot) \, m(\cdot) \,\equiv\,0\,,~~~~~
\tau^m_{\pi \pi}(\cdot)\,=\,   \pi' (\cdot)\,\tau^m(\cdot)\,   \pi (\cdot)  \,=\, \tau_{mm}^{\pi}(\cdot)\,, ~~~~~ \tau^m_{~ \widetilde \pi^{(p)}  \widetilde \pi^{(p)}}(\cdot)\,=\,p^2 \cdot \tau^m_{\pi \pi}(\cdot)
\,.
\end{equation}

\medskip
We shall take $\, m(\cdot) \equiv \mu (\cdot)$ from now on. The relative performance of $\, \pi(\cdot)$ with respect to $\, \mu(\cdot)$ is given in (1.2.16) of Fernholz (2002) by
\begin{equation}
\label{B.5}
d \, \log \left( \frac{Z^{\pi} (t)}{Z^{\mu} (t)} \right)
\,=\, \sum_{i=1}^n   (\pi_i (t) - \mu_i (t)) \, d \log \mu_i (t) \,+\,
\left( \gamma^{\pi}_* (t) -  \gamma^{\mu}_* (t) \right) \, dt\,.
\end{equation}
Writing this expression for $\, \widetilde \pi^{(p)} (\cdot)$  in place of $\, \pi(\cdot)$, recalling $\,  \widetilde \pi^{(p)} -\mu = p (\pi-\mu)$ from (\ref{B.1}), and then subtracting (\ref{B.5}) multiplied by $\,p\,$, we obtain
 \begin{equation}
\label{B.6}
d \, \log \left( \frac{Z^{\widetilde \pi^{(p)}} (t)}{Z^{\mu} (t)} \right)
\,=\, p \cdot
d \, \log \left( \frac{Z^{\pi} (t)}{Z^{\mu} (t)} \right)\,+\,
(p-1) \, \gamma^{\mu}_* (t)   \, dt\,+\,
\left( \gamma^{\widetilde \pi^{(p)}}_* (t) -p \, \gamma^{\pi}_* (t) \right) \, dt\,.
\end{equation}
But now recall the expressions of  (\ref{NI}), (\ref{B.4}) and (5.4), to obtain
\[
2 \, \left( \gamma^{\widetilde \pi^{(p)}}_* (t) -p  \gamma^{\pi}_* (t) \right)
 \, = \,
\sum_{i=1}^n \left(  \widetilde \pi^{(p)}_i (t) - p\, \pi_i (t) \right)   \tau^{\mu}_{ii}(t)\, -\, \tau^{\mu}_{~ \widetilde \pi^{(p)}  \widetilde \pi^{(p)}}( t)\, +\, p\, \tau^{\mu}_{\pi \pi} (t)~~~~~~~~~~~~~~~
 \]
\[
~~~~~~~ = \,
(1-p) \cdot \sum_{i=1}^n  \,
\mu_i(t) \,\tau^{\mu}_{ii}(t)
\, +\, p \, \tau^{\mu}_{\pi \pi} (t)
\, -\,p^2 \tau^{\mu}_{\pi \pi} (t)
\, =\,  (1-p) \cdot \left[\,
2\, \gamma^{\mu}_* (t) \,+\,
p \, \tau^{\mu}_{\pi \pi} (t)\, \right]\,.
\]
Substituting back into (\ref{B.6}) we get
 \begin{equation}
\label{B.7}
  \log \left( \frac{Z^{\widetilde \pi^{(p)}} (T)}{Z^{\mu} (T)} \right)
\,=\, p \cdot
  \log \left( \frac{Z^{\pi} (T)}{Z^{\mu} (T)} \right)\,+\, \frac{
\,p(1-p)\,}{2} \, \int_0^T
\tau^{\mu}_{\pi \pi}(t)   \, dt
\end{equation}
and note that the last term is  non-negative, by (\ref{B.3}).

\medskip
\noindent
{\bf 8.1 Lemma:} ~{\it Suppose that the extended portfolio $\, \pi (\cdot)$ is such that the conditions
\begin{equation}
\label{B.8}
P  \left(  \frac{Z^{\pi} (T)}{Z^{\mu} (T)} \ge \beta \right) =1 ~~~~~~\hbox{or}~~~~~~~~
P  \left(  \frac{Z^{\pi} (T)}{Z^{\mu} (T)} \le \frac{1}{\beta} \right) =1
\end{equation}
and
\begin{equation}
\label{B.9}
P  \left(  \int_0^T
\tau^{\mu}_{\pi \pi}(t)   \, dt\, \ge \, \eta \right) =1
\end{equation}
hold, for some
$\, \beta >0$ and $\, \eta >0$. Then there exists an extended portfolio $\, \widehat \pi (\cdot)$ such that}
\begin{equation}
\label{B.10}
P \left( Z^{\widehat \pi} (T) < Z^{\mu} (T) \right) \,=\,1\,.
\end{equation}

\smallskip
\noindent
{\sl  8.1  Remark:}~ Condition (\ref{B.8}) postulates that the extended portfolio $\, \pi(\cdot)$ is ``not very different" from the market portfolio. But condition (\ref{B.9}) mandates that  $\, \pi(\cdot)$ ``must be sufficiently  different" from the market portfolio; indeed,   $\, \int_0^T
\tau^{\mu}_{\pi \pi}(t)   \, dt\, \ge \, \eps\, \sum_{i=1}^n \int_0^T |\pi_i (t) - \mu_i (t) |^2\, dt\,$ from (\ref{B.3}), so (\ref{B.9}) holds if the expression $\,  || \pi - \mu ||_{{\bf L}^2 ([0,T])} \,$ is bounded away from zero, a.s.

\medskip
\noindent
{\it Proof of Lemma 8.1:}~ If we have $\, P \left[\, \left(   Z^{\pi} (T)/Z^{\mu} (T)  \right) \le 1 / \beta \,\right] =1\,$, then it suffices to take $\, p> 1+ (2/ \eta) \cdot \log (1 / \beta)\,$ and observe from  (\ref{B.9}), (\ref{B.7}) that $\, \widehat \pi(\cdot) \equiv \widetilde \pi^{(p)} (\cdot)\,$ satisfies
\[
\log \left( \frac{Z^{\widehat \pi} (T)}{Z^{\mu} (T)} \right)\,\le\,
p \cdot \left[ \, \log \left( \frac{1}{\beta}  \right) + \frac{ \eta}{2}\, (1-p) \, \right] \,<\,0\,,~~~\hbox{a.s.}
\]
If  on the other hand we have $\, P \left[\, \left(   Z^{\pi} (T)/Z^{\mu} (T)  \right) \ge  \beta \,\right] =1\,$, then it suffices to take $\, p < \min \bigl( 0\,,\,  1- (2/ \eta) \cdot \log (1 / \beta) \bigr)\,$ and observe from (\ref{B.7}) that $\, \widehat \pi (\cdot) \equiv \widetilde \pi^{(p)} (\cdot)\,$ satisfies
\[
\log \left( \frac{Z^{\widehat \pi} (T)}{Z^{\mu} (T)} \right)\,\le\,
p \cdot \left[ \, - \log \left( \frac{1}{\beta}  \right) + \frac{ \eta}{2}\, (1-p) \, \right] \,<\,0\,,~~~\hbox{a.s.}~~~~~~~~~~~~\diamond
\]

\medskip
\noindent
{\bf 8.1 Example:} ~ With $\, \pi = e_1 = (1, 0 , \cdots, 0)'\,$ and $\, m(\cdot) \equiv  \mu(\cdot) \,$ the market portfolio, take $\, p>1$ to be detemined  in a moment, and define the extended portfolio
\begin{equation}
\label{B.11}
 \, \widehat \pi (t) \,:=\, \widetilde \pi^{(p)} (t)\,=\,
p\, e_1\,+\, (1-p)\, \mu(t)\,,~~~~~ 0 \le t < \infty\,.
\end{equation}
which takes a long position in the first stock and a short position in the market. (This is not a very easy portfolio to implement in actual practice.) In particular, $\, \widehat \pi_1 (t) \,=\,
p \,+\, (1-p)\, \mu_1 (t)\,$   and $\, \widehat \pi_i (t) \,=\,
  (1-p)\, \mu_i (t)\,$ for $\,i=2, \cdots, n\,$. Then we have
 \begin{equation}
\label{B.12}
  \log \left( \frac{Z^{\widehat \pi} (T)}{Z^{\mu} (T)} \right)
\,=\, p \cdot \left[\,
 \log \left( \frac{\mu_1 (T)}{\mu_1 (0)} \right)\,-\, \frac{
\,p-1\,}{2} \, \int_0^T
\tau^{\mu}_{1 1}(t)   \, dt\,\right]\,
\end{equation}
from (\ref{B.7}). But taking $\, \beta := \mu_1 (0)\,$ we have $\, ( \mu_1 (T) / \mu_1 (0) ) \le 1 / \beta\,$, and if the market is weakly diverse on $[0,T]$ we obtain from (5.10) and the Cauchy-Schwarz inequality
\begin{equation}
\label{B.13}
\int_0^T
\tau^{\mu}_{1 1}(t)   \, dt\, \ge \,
\eps \,  \int_0^T
\bigl( 1 - \mu_{(1)} \bigr)^2 \,dt
 \, > \,  \eps \delta^2 T  \, =:  \, \eta   \,.
\end{equation}
>From  Lemma 8.1   the market portfolio represents then an arbitrage opportunity with respect to the extended portfolio $\, \widehat  \pi (\cdot)\,$ of (\ref{B.11}), provided that for any given $\,T \in (0,\infty)\,$ we select $\, p \,>\, p(T) := 1 \,+\, \frac{2}{\,  \eps \delta^2 T  \, } \cdot \log \left( \frac{1}{\mu_1 (0)} \right)\,$.  Note that $\, \lim_{T \downarrow 0}  p(T) = \infty\,$. ~~~~~~~~~~~~$\diamond$

\bigskip
The extended portfolio $\widehat \pi (\cdot)$ of (\ref{B.11}) can be used to create {\sl all-long} portfolios that underperform (Example 8.2) or outperform (Example 8.3) the market portfolio $\mu(\cdot)$, {\sl over any given time-horizon} $T\in (0,\infty)$. The idea is to ``embed $\widehat \pi (\cdot)$  in a sea of market portfolio, swamping the short positions while retaining the essential portfolio characteristics". Crucial in these constructions is the a.s. comparison
\begin{equation}
\label{B.14}
Z^{\widehat \pi} (t) \, \le \,
\left( \frac{\mu_1 (t)}{\mu_1 (0)} \right)^p \cdot
Z^{\mu} (t)\,,~~~~ 0 \le t < \infty\,,
\end{equation}
a direct consequence of (\ref{B.12}). Here and in what follows we assume $\, Z^{ \mu} (0) =Z^{\widehat \pi} (0)=1$.

\medskip
\noindent
{\bf 8.2 Example:} ~ Consider an investment strategy $\rho(\cdot)$ that places one dollar in the portfolio $\widehat \pi (\cdot)$ of (\ref{B.11})  and $\, (p-1)/( \mu_1 (0))^p\,$ dollars in the market portfolio $\mu(\cdot)$ at time $t=0$, and makes no change afterwards. The number $p$ is chosen as in Example 8.1. The value $\, Z^{ \rho} (\cdot)$ of this strategy is clearly
$\,
Z^{ \rho} (t)\,=\, Z^{\widehat \pi}(t) +
\frac{p-1}{\left( \mu_1 (0) \right)^p} \cdot Z^{ \mu} (t) \,>\,0$, $\,
 0 \le t < \infty
$, and is generated by the extended portfolio with weights
$$
\rho_i (t) \,=\, \frac{1}{Z^{ \rho} (t)} \, \left[\,
\widehat \pi_i (t) \cdot Z^{\widehat \pi}(t) +
\frac{p-1}{\left( \mu_1 (0) \right)^p} \cdot \mu_i (t)\,Z^{ \mu} (t)\,
\right]\,, ~~~\hbox{for}~~ i = 1, \cdots, n\,.
$$
Clearly $\, \sum_{i=1}^n \rho_i (t)=1$; and since both $\widehat \pi_1 (t)$ and $\mu_1 (t)$ are positive, we have $\rho_1 (t) >0$ as well. To check that $\rho (\cdot)$ is an all-long portfolio, observe that the dollar amount invested by it at time $t$ in any stock $\,i=2, \cdots, n\,$, is
$$
- (p-1) \, \mu_i (t) \cdot Z^{\widehat \pi}(t) +
\frac{p-1}{\left( \mu_1 (0) \right)^p} \cdot \mu_i (t)\,Z^{ \mu} (t)\,
\ge \,
\frac{(p-1)\mu_i (t)}{\left( \mu_1 (0) \right)^p}\,
\bigl[\, 1 - \left( \mu_1 (t) \right)^p\, \bigr] \, Z^{ \mu} (t)\,
>\,0
$$
thanks to (\ref{B.14}). On the other hand, $\rho(\cdot)$ {\it underperforms} at $t=T$ a market portfolio that starts out with the same initial capital $\, z := Z^{ \rho} (0) = 1 + (p-1)/( \mu_1 (0))^p\,$, since $\rho(\cdot)$ holds a mix of $\mu(\cdot)$ and $\widehat \pi (\cdot)$, and $\widehat \pi (\cdot)$ underperforms the market at $t=T$:
$$
Z^{ \rho} (T) \,=\, Z^{\widehat \pi}(T) +\frac{p-1}{\left( \mu_1 (0) \right)^p} \,Z^{ \mu}(T)\,<\, z Z^{ \mu}(T) = Z^{ \,z,\mu}(T)~~~\hbox{a.s., from (8.10)}.
$$

\medskip
\noindent
{\bf 8.3 Example:} ~ Now consider a  strategy $\eta(\cdot)$ that invests $\, p/( \mu_1 (0))^p\,$ dollars in the market portfolio and $\,-1$ dollar in   $\widehat \pi (\cdot)$   at time $t=0$, and makes no change thereafter. The number $p>1$ is chosen again as in Example 8.1. The value $\, Z^{ \eta} (\cdot)$ of this strategy is
\begin{equation}
\label{B.15}
Z^{ \eta} (t)\,=\,
\frac{p}{\left( \mu_1 (0) \right)^p} \cdot Z^{ \mu} (t)
- Z^{\widehat \pi}(t) \,\ge\,
\frac{Z^{ \mu} (t) }{\left( \mu_1 (0) \right)^p}\,
\bigl[\, p - \left( \mu_1 (t) \right)^p\,\bigr] \,>\,
0\,,~~~~ 0 \le t < \infty\,,
\end{equation}
thanks to (\ref{B.14}) and $p>1>\left( \mu_1 (t) \right)^p$. As before, $Z^{ \eta} (\cdot)$  is generated by an extended portfolio $\eta(\cdot)$ with weights
\begin{equation}
\label{B.16}
\eta_i (t) \,=\, \frac{1}{Z^{ \eta} (t)} \, \left[\,
\frac{p }{\left( \mu_1 (0) \right)^p} \cdot \mu_i (t)\,Z^{ \mu} (t)-
\widehat \pi_i (t) \cdot Z^{\widehat \pi}(t)
\,\right]\,, ~~~  ~ i = 1, \cdots, n\,
\end{equation}
that satisfy $\sum_{i=1}^n \eta_i (t)=1$. Now for $\,i=2, \cdots, n\,$ we have $\, \widehat \pi_i (t) = - (p-1)\,  \mu_i (t) <0$, so $\, \eta_2(\cdot), \ldots, \eta_n(\cdot)$ are strictly positive.  To check that $\eta (\cdot)$ is an all-long portfolio, it remains to verify $\eta_1 (t) \ge 0$; but the dollar amount
$$
\frac{p\mu_1 (t) }{\left( \mu_1 (0) \right)^p} \cdot Z^{ \mu} (t)-
\bigl[\, p - (p-1) \mu_1 (t)\, \bigr]\cdot Z^{\widehat \pi}(t)
$$
invested by $\eta(\cdot)$  in the first stock at time $t$, dominates
$\,
\frac{p\mu_1 (t) }{\left( \mu_1 (0) \right)^p} \cdot Z^{ \mu} (t)-
\bigl[ p - (p-1) \mu_1 (t) \bigr]\cdot
\left( \frac{\mu_1 (t)}{\mu_1 (0)} \right)^p Z^{\mu}(t)
\, $, or equivalently the quantity
$$
\frac{Z^{\mu}(t) \mu_1 (t) }{\left( \mu_1 (0) \right)^p} \cdot
\left[\,
(p-1) \, \left( \mu_1 (t) \right)^p + p \, \left\{
1 - \left( \mu_1 (t) \right)^{p-1} \right\} \, \right] \,>\,0\,,
$$
again thanks to (\ref{B.14}) and $p>1>\left( \mu_1 (t) \right)^p$. Thus $\eta(\cdot)$ is indeed an all-long portfolio.

On the other hand, $\eta(\cdot)$ {\it outperforms} at $t=T$ a market portfolio  with the same initial capital of $\, \zeta := Z^{ \eta} (0) = p/( \mu_1 (0))^p-1>0\,$ dollars, because $\eta(\cdot)$ is long in  the market $\mu(\cdot)$ and short in the extended portfolio $\widehat \pi (\cdot)$, which  underperforms the market at $t=T$:
$$
Z^{ \eta} (T) \,=\, \frac{p}{\left( \mu_1 (0) \right)^p}\, Z^{ \mu}(T)
- Z^{\widehat \pi}(T) \,>\, \zeta Z^{ \mu}(T) = Z^{ \,\zeta,\mu}(T)~~~\hbox{a.s., from (8.10)}.
$$

\medskip


\section{Hedging
in Weakly Diverse Markets}

\setcounter{equation}{0}
\setcounter{Assumption}{0}
\setcounter{Theorem}{0}
\setcounter{Proposition}{0}
\setcounter{Corollary}{0}
\setcounter{Lemma}{0}
\setcounter{Remark}{0}
\setcounter{Example}{0}

Suppose now that we place a small investor in a   market $\, {\cal M}$ as in (\ref{2.1})-(\ref{2.5}) but allow him to invest also in a money-market with interest rate $\, r: [0,\infty) \times \Omega \rightarrow [0,\infty)$: a progressively measurable and locally integrable process. A dollar invested at time $t=0$ in the money market grows to $\, B(T) = \exp \{ \int_0^T r(u)\,du\}\,$ at time $t=T$.

Starting with initial capital $ z>0$, the investor can choose at any time $t$ a {\bf trading strategy} $\, \f (t) = ( \f_1 (t) , \cdots, \f_n (t) )'\,$. With $\, Z^{z, \f}( t)$  denoting the value  of the   strategy at time $t$, the quantity  $\, \f_i (t)$ is the dollar amount invested in the $i^{th}$ stock   and $\,Z^{z, \f}( t) - \sum_{i=1}^n \f_i (t)$ the amount held in the money-market. These quantities are real-valued, any one of them may be negative: selling stock  short is allowed, as is borrowing from (as opposed to depositing into) the money-market. We require only that the trading strategy $\, \f (\cdot)\,$ be progressively measurable and satisfy
 $\, \sum_{i=1}^n \int_0^T [ \,(\f_i (t))^2 + | \f_i (t)| | b_i (t) - r(t)| \,] \,dt < \infty\,$ a.s., on any given time-horizon $\, [0,T]$. With this understanding, the value-process $\, Z(\cdot) \equiv Z^{z, \f}( \cdot)$ satisfies
\begin{equation}
\label{C.1}
dZ ( t)\,
  = \, \sum_{i=1}^n \f_i (t)   \cdot \frac{d X_i (t)}{X_i (t)} \,+\, \left( Z ( t)  - \sum_{i=1}^n \f_i (t) \right) \cdot
\frac{dB( t)}{B(t)}~~~~~~~~~~~~~~~~~~~~~~~~~~~~~~~~~~~~~~~
\end{equation}
\[
~~  =\,r(t) Z ( t) dt\,+\,  \sum_{i=1}^n \f_i (t)
\left( ( b_i (t) - r(t) ) \, dt + \sum_{\nu=1}^m \sigma_{i \nu} (t) \, dW_{\nu} (t) \right)= r(t)Z ( t) dt\,+\, \f^{\prime} (t)   \sigma (t)   d \widehat W (t)\,,
\]
a simple linear equation. We have introduced the processes
\begin{equation}
\label{C.2}
\widehat W (t)\,:=\, W(t) + \int_0^t \vartheta (s) \, ds\,,~~~~~~
\vartheta (t)\,:=\, \sigma^{\prime} (t) \bigl( \sigma  (t)  \sigma^{\prime} (t)  \bigr)^{-1} \, [  b(t) - r(t) \hbox{{\bf 1}} ]
\end{equation}
with $\,${\bf 1}$\,\,=\, (1, \cdots , 1)' \in \R^n$, in terms of which we can write the equation (\ref{2.1}) in the form
\begin{equation}
\label{2.1'}
 d \left( \frac{X_i (t)}{B(t)} \right) \, =\,  \left( \frac{X_i (t)}{B(t)} \right) \cdot \sum_{\nu=1}^m \sigma_{i \nu} (t) \, d \widehat W_{\nu} (t)\,,~~~~~~~i =1, \cdots, n\,.
\end{equation}

\medskip
 The solution of the equation (\ref{C.1}) is given by
\begin{equation}
\label{C.4}
\frac{Z^{z, \f}( t)}{B(t)} \,=\, z\, +\,  \int_0^t
\frac{\f^{\prime} (s) }{B(s)}\, \sigma (s) \, d \widehat W (s)  \,, \quad 0 \le t < \infty\,.
\end{equation}
We  shall denote by $\, \Phi_T (z)$ the class of   trading strategies $\f (\cdot)\,$ that satisfy $\, P \bigl[\, Z^{z, \f}( t) \ge 0\,,$ $ \forall \,0 \le t \le T\,\bigr]=1\,$ for a given $\, T \in (0,\infty)$, and set $\, \Phi (z) := \cap_{0 < T <\infty}  \Phi_T (z)\,$. This class contains the extended portfolios of section 8: if $\pi(\cdot)$ is an extended portfolio and $\, Z^{\pi}(\cdot)$ its value-process with initial capital $\, Z^{\pi}(0) = z>0$, then $\, \f_i (\cdot) := \pi_i (\cdot) Z^{\pi}(\cdot)$, $1 \le i \le n$ defines a trading strategy, and $\, Z^{z, \f}( \cdot) \equiv Z^{\pi}(\cdot)>0\,$ satisfies the analogue $\, d (Z^{\pi} (t)/ B(t)) = (Z^{\pi} (t)/ B(t)) \cdot \pi' (t) \sigma (t) d \widehat W (t)\,$ of (\ref{C.4}).

\smallskip
\noindent
{\sl 9.1 Remark:} If $\, {\cal M}$ is weakly diverse on some finite horizon $[0,T]$, then the process
\begin{equation}
\label{C.5}
L( t) \,:=\,   \exp \left(  - \int_0^t
\vartheta^{\prime} (s)  \, d  W (s) \,-\,
\frac{1}{2} \int_0^t ||
\vartheta (s)   ||^2\, d s \right) \, , ~~~~~ 0 \le t < \infty
\end{equation}
is a local martingale and a supermartingale but {\it is not a martingale}. For if it were, then the measure $\,Q_T (A) := E [ L(T)\cdot 1_A]$ would be a probability on $ \F (T)$. Under this probability measure, the process $\widehat W (\cdot)$ of (\ref{C.2}) would be Brownian motion  and the discounted price-processes $\, X_i(\cdot)/ B(\cdot)\,$ would be martingales on the interval $[0,T]$, from (\ref{2.1'}), (\ref{2.3}). But this would proscribe  (\ref{AO})
on this interval for any two extended portfolios $\, \pi(\cdot)$ and $\, \rho (\cdot)$, contradicting (\ref{4.5}) (see  Appendix A for a formal argument along these lines).

Thus, in a weakly diverse market the process $\, L(\cdot)$ of (\ref{C.5}) is a {\sl strict local martingale} in the sense of Elworthy et al. (1997): we have $\, E[L(t)] <1\,$ for every $\, t \in (0, \infty)$.

\medskip
\noindent
{\sl 9.2 Remark:}  Because $\, L(\cdot)$ is a local martingale there exists an increasing sequence $\, \{ S_k \}_{k \in {\bf N}}\,$ of stopping times with $\,\lim_{k \rightarrow \infty} S_k = \infty\,$ a.s.  such that $\, L( \cdot \wedge S_k)\,$ is a martingale for every $\, k \in {\bf N}$; for instance, take $\, S_k = \inf \{ t \ge 0\,|\, \int_0^t ||\vartheta (s) ||^2 \, ds \ge k\, \}$. Thus, if we replace $T$  by $T\wedge S_k$ in (\ref{AO}), this property  cannot hold for {\it any} extended portfolios $\, \pi(\cdot)$ and $\, \rho (\cdot)$: {\sl there is no  possibility for relative arbitrage on the horizon $\,[0, T \wedge S_k]\,$ for any} $k \in {\bf N}$. But in the limit as $\, k \rightarrow \infty$ a relative arbitrage of the type (\ref{AO}) appears  as in (4.5) or Example 8.1, if $\, {\cal M}$ is weakly diverse on  $[0,T]$.

\medskip
\noindent  $\bullet$ ~
The failure of the exponential process $\, L(\cdot)$ in (\ref{C.5}) to be a martingale does not preclude, however, the possibility for hedging contingent claims in a market ${\cal M}$ which is weakly diverse on some finite horizon $[0,T]$. To see why,   consider   an $\F (T)-$measurable random variable $\, Y: \Omega  \rightarrow [0, \infty)$ that satisfies
\begin{equation}
\label{C.3}
0 <  y_0 := E [\, Y L(T) / B(T)\,] < \infty\,.
\end{equation}
 If we view $Y$ as a liability (contingent claim) that the investor faces and has to  cover (hedge) at time $t=T$, the question is to characterize the smallest amount of initial capital that allows the investor to hedge this liability without risk; namely, the {\it hedging price}
\begin{equation}
\label{C.6}
h \,:=\, \inf \{ z > 0\,|\, \,\hbox{there exists  } \f (\cdot) \in \Phi_T (z) \,\hbox{  such that } Z^{z, \f} (T) \ge Y ~~\hbox{holds a.s.} \}\,.
\end{equation}

We proceed as in the standard treatment of this question (e.g. Karatzas \& Shreve (1998), Chapter 2) but under the probability measure $P$, the only one now at our disposal: from (\ref{C.1})-(\ref{2.1'}) and the differential equation $\, dL (t) = - L(t) \, \vartheta^{\prime} (t) \, dW(t)\,$ for the exponential process $\, L(\cdot)\,$ of (\ref{C.5}), we obtain that each of the processes
\begin{equation}
\label{2.1"}
\widetilde X_i (t) :=
\frac{L(t) X_i (t)}{X_i (0) B(t)} \,=\, 1 + \int_0^t  \widetilde X_i  (s)   \cdot \sum_{\nu=1}^{m}
( \sigma_{i \nu} (s) - \vartheta_{\nu} (s) ) \, dW_{\nu} (s)\,,~~~i = 1, \cdots, n
\end{equation}
\begin{equation}
\label{C.7}
\widetilde Z^{ \f} (t) :=
\frac{L(t) Z^{z, \f} (t)}{z B(t)}\,=\, 1 + \int_0^t  \frac{L(s)  }{z B(s)}\, \Bigl(  \f^{\prime} (s) \sigma (s) - Z^{z, \f} (s) \vartheta^{\prime} (s) \Bigl)\, dW(s)
\end{equation}
(products of $L(\cdot)$ with the discounted stock-prices and
with the discounted values of investment strategies in $\Phi (z)$, respectively) is a non-negative local martingale, hence a supermartingale. It is not hard to see (in Appendix A) that
\begin{equation}
\label{AO3}
\hbox{the processes}~~~ \widetilde X_i (\cdot) \,, ~ i = 1, \cdots, n ~~\hbox{of (\ref{2.1"}) are strict local  martingales}\,.
\end{equation}
In  particular, $ \, E [ \,  L(T) X_i (T) / B(T)\, ] < X_i (0)\,$   holds for all $\, T \in (0, \infty)\,$.

For any $z>0$ in the set of (\ref{C.6}), there exists some $\,  \f (\cdot)  \in \Phi_T (z) \, $ such that
\begin{equation}
\label{C.10}
E [ \,Y L(T) / B(T)\, ] \,\le\,  E [\, Z^{z, \f} (T)  L(T) / B(T)\,  ]\,\le\, z\,,
\end{equation}
and so  $\, y_0 = E [ Y L(T)/ B(T) ] \le h\,$.

\medskip
\noindent  $\bullet$ ~
Let us suppose from now on that $m=n$, i.e., that we have exactly as many sources of randomness as there are stocks in the market ${\cal M}$; that the square-matrix $\sigma (t, \omega) = \{ \sigma_{ij} (t, \omega) \}_{1 \le i,j \le n}\,$ is invertible for every $\,(t, \omega) \in [0,T] \times \Omega$; and that the filtration ${\bf F} = \{ \F(t) \}_{0 \le t \le T}\,$ is generated by the Brownian Motion $W(\cdot)$ itself, namely,  $\F(t) = \sigma (W(s); ~ 0 \le s \le t)\,$. The martingale representation property of this Brownian filtration gives
\begin{equation}
\label{C.8}
M(t) \,:=\, E \, \left[ \, \frac{Y L(T)}{B(T)}\,   \Big \vert \,  \F(t) \, \right] \,=\, y_0 + \int_0^t \psi^{\prime} (s) \, dW(s) \, \ge \, 0 \,,~~~~~~ 0 \le t \le T
\end{equation}
for some progressively measurable process $\, \psi : [0,T] \times \Omega  \rightarrow \R^n\, $ with $\, \sum_{i=1}^n  \int_0^T ( \psi_i (t) )^2\, dt < \infty\,$ a.s. Setting $\, \widehat Z (\cdot) := M(\cdot) B (\cdot)  / L(\cdot)\,$, $\, \widehat \f (\cdot) := \frac{B (\cdot)}{L(\cdot) } \Bigl( \sigma^{-1} (\cdot) \Bigr)^{\prime} \bigl(  \psi(\cdot)  +M(\cdot) \vartheta (\cdot) \bigr)\,$ and comparing  (\ref{C.7}) with (\ref{C.8}), we observe $\, \widehat Z (0) = y_0\,$,  $\, \widehat Z (T) = Y\,$ and $\, \widehat Z (\cdot) \equiv Z^{y_0, \widehat \f} (\cdot) \ge 0\,$,  almost surely.

Therefore, the  trading strategy $\, \widehat \f (\cdot)$ is in $\, \Phi_T (y_0)\,$ and satisfies $~Z^{\,y_0, \widehat \f} (T) = Y\,$ a.s. This means that $y_0$  belongs to the set on the right-hand-side of (\ref{C.6}), and so $\, y_0 \ge h\,$. But we have already established the reverse inequality (actually in much greater generality), so for the hedging price of (\ref{C.6}) we get the {\sl Black-Scholes-type formula}
\begin{equation}
\label{C.9}
h \,=\, E [\, Y L(T) / B(T)\,]
\end{equation}
under the assumptions of the preceding paragraph. In particular, a market ${\cal M}$ that is weakly diverse  --  hence without an equivalent probability measure under which discounted stock-prices are (at least local) martingales -- can  nevertheless be {\bf complete}.


\smallskip
\noindent
{\sl 9.3 Remark:}  In terms of Delbaen \& Schachermayer (1998) we are now in a situation where {\it ``no arbitrage"}  holds for trading strategies with non-negative value (in the sense that  $ \f (\cdot) \in \Phi_T (0)\,$ implies $\, Z^{0, \varphi} (T) = 0\,$ a.s. in (\ref{C.10})), where    $L(\cdot)$, $\widetilde X_i (\cdot)$ and $\widetilde Z^{\f} (\cdot)$ are   local martingales, but where  {\it ``free lunch with vanishing risk"} also exists (as illustrated in Remark 9.2; a related notion was introduced and studied in Lowenstein \& Willard (2000)  under their terminology ``cheap thrill"). We owe this observation to Profs.  Steven Shreve and Julien Hugonnier.


\medskip
\noindent
{\sl 9.1 EXAMPLE:  A European Call-Option.}
Consider the contingent claim $\, Y = \bigl( X_1 (T) - q \bigr)^+\,$: this is a European call-option with strike-price $q>0$ on the first stock.
Let us assume also that the interest-rate process $r(\cdot)$ is bounded away from zero, namely that $\, P [\, r(t) \ge r \,,~ \forall \, t \ge 0\,]\,=\,1\,$ holds for some $r>0$, and that the market ${\cal M}$ is weakly diverse on all time-horizons $T \in (0,\infty)$ sufficiently large. Then for the hedging price  of this contingent claim,  written now as a function $\,h(T)\,$ of the time-horizon, we have from  (\ref{AO3}),  (\ref{C.9}) and $\, E [L(T)]<1$:
\begin{eqnarray*}
X_1 (0) \,
&> &\, E [ \,L(T) X_1(T) / B(T)\,]
 \ge \,
 E [ \,L(T) ( X_1(T)-q)^+ / B(T)\,] \,=\, h(T)
 \\
& \ge & \,
E [ \,L(T) X_1(T) / B(T)\,] \,-\, q \cdot E \left( L(T) \cdot e^{\, - \int_0^T r(t)\,dt} \right)
\\
& \ge & \,
E [ \,L(T) X_1(T) / B(T)\,] \,-\, q\,  e^{\, - rT  }  \, E [ L(T)]
\, > \,
E [ \,L(T) X_1(T) / B(T)\,] \,-\, q\,  e^{\, - rT  }  \,,
\end{eqnarray*}
because $\, L(\cdot) X_1 (\cdot) / B(\cdot)\,$ is a supermartingale and a strict local martingale. Therefore
\begin{equation}
\label{C.11}
0 \le h (\infty) :=
\lim_{T \rightarrow \infty}\, h(T) \, =\,
\lim_{T \rightarrow \infty} \downarrow \, E \left( \frac{L(T) X_1 (T)}{B(T)} \right)
\, < \,  X_1 (0)\,:
\end{equation}
the value of the option is {\sl strictly less} than the price of the underlying stock at time $t=0$, and  tends to the value $h (\infty) \in [0, X_1 (0))\, $ as the horizon increases without limit.

We claim that if $\, {\cal M}$ is uniformly weakly diverse over some $\, [T_0, \infty)$, then the limit in (\ref{C.11}) is actually zero: {\sl a European call that can never be exercised is worthless}. Indeed, for every fixed $p \in (0,1)$ and $T \ge \frac{2 \log n}{p \varepsilon \delta} \vee T_0\,$, the quantity
\[
E \left( \frac{L(T) }{B(T)} X_1 (T)\right) \le
E \left( \frac{L(T) }{B(T)} Z^{\mu} (T)\right) \le
E \left( \frac{L(T) }{B(T)} Z^{\pi^{(p)}} (T)\right) \cdot
n^{\frac{1-p}{p}} e^{\, - \varepsilon \delta (1-p) T / 2}
\]
is dominated by  $\, Z(0) \,
n^{\frac{1-p}{p}} \cdot e^{\, - \varepsilon \delta (1-p) T / 2}
 \,$, from (3.1), (11.4) and the supermartingale property of $\, L(\cdot) Z^{\pi^{(p)}} (\cdot) / B(\cdot)\,$. Letting $\,T \rightarrow \infty\,$ as in (\ref{C.11}), this leads to the claim $\, h(\infty) =0\,$.

\smallskip
\noindent
{\sl 9.4 Remark:}   Note the sharp difference between this case and the situation where an equivalent martingale measure exists on any finite time-horizon;  namely, when both $\,L(\cdot)\,$ and $\,L(\cdot) X_1 (\cdot) / B(\cdot)\,$ are martingales. Then $\, E [ L(T) X_1 (T) / B(T)] = X_1 (0)\,$ holds for all $ T \in (0, \infty)$, and   $\, h(\infty)= X_1 (0)$: as the time-horizon increases without limit, the hedging price of the call-option approaches  the current stock price (Karatzas \& Shreve (1998), page 62).

\medskip
\noindent
{\sl 9.2 EXAMPLE:  Put-Call Parity.} ~ Suppose that $\, \Xi_1 (\cdot), \, \Xi_2 (\cdot)\,$ are positive, continuous and adapted processes, representing the values of two different assets in a market $\, {\cal M}\,$ with $\, r(\cdot) \equiv 0$. Let us set
\[
Y_1 \,:=\, \Bigl( \Xi_1 (T) - \Xi_2 (T) \Bigr)^+
~~~~~~~\hbox{and}~~~~~~~
Y_2 \,:=\, \Bigl( \Xi_2 (T)-\Xi_1 (T) \Bigr)^+\,;
\]
then from (\ref{C.9}) the quantity $\, h_j = E [ L(T) Y_j]\,$ is the hedging price at time $t=0$ of a contract that offers its holder the right, but not the obligation, to exchange asset 2 for asset 1 with $j=1$ (resp., asset 1 for asset 2 with $j=2$) at time $t=T$. We have clearly
\[
h_1 - h_2 \,=\, E[\, L(T) \, (  \Xi_1 (T) - \Xi_2 (T) )\,]\,,
\]
and  say that the two assets are in {\sl put-call parity}  if $\,h_1 - h_2 \,=\, \Xi_1 (0) - \Xi_2 (0) \,$. This will be the case when both  $\, L(\cdot)\, \Xi_1 (\cdot)$, $ \, L(\cdot) \,\Xi_2 (\cdot)\,$ are martingales. (For instance, whenever (4.6) is valid we can take $\, \Xi_j (\cdot) \equiv X_i (\cdot)\,$ or $\, \Xi_j (\cdot) \equiv Z^{\pi} (\cdot)\,$ for any $i=1, \cdots, n$, $j=1,2$ and any extended portfolio $\, \pi (\cdot)\,$; then put-call parity holds as in  Karatzas \& Shreve (1998), p.50.)

It is easy to see that {\it put-call parity need not hold if ${\cal M}$ is weakly diverse:} for instance, take $\, \Xi_1 (\cdot) \equiv Z^{\, \mu} (\cdot)\,$, $\, \Xi_2 (\cdot) \equiv Z^{\, \widehat \pi} (\cdot)\,$
 with $\,  Z^{\, \mu} (0)=Z^{\, \widehat \pi} (0)\,$ in the notation of (3.1) and (8.11), and observe from (8.10) that $\, h_1 - h_2 \,=\, E[\, L(T) \, (  Z^{\, \mu} (T) - Z^{\, \widehat \pi} (T) )\,] \,>\, 0\,=\,
Z^{\, \mu} (0)-Z^{\, \widehat \pi} (0)\,$.

\section{Concluding Remarks}

\setcounter{equation}{0}
\setcounter{Assumption}{0}
\setcounter{Theorem}{0}
\setcounter{Proposition}{0}
\setcounter{Corollary}{0}
\setcounter{Lemma}{0}
\setcounter{Remark}{0}
\setcounter{Example}{0}

We have presented  examples of diverse and weakly diverse market models posited in Fernholz (1999, 2002), and   shown that the ``diversity-weighted" portfolio of (\ref{4.4}) represents an arbitrage opportunity relative to a weakly-diverse  market over a sufficiently long time-horizon. We have also shown that weakly-diverse markets are themselves arbitrage opportunities relative to suitable extended portfolios,  over arbitratry time-horizons; in particular, no equivalent martingale measure can exist for such markets. But we have also shown that,  even in the context of a diverse market, this
does not in any way interfere with the development of option pricing; quite the contrary, one is led to more realistic values for warrants over exceedingly long time-horizons. A similar treatment is possible for utility maximization problems, along the lines of Karatzas et al. (1991). It would also be of interest to determine the optimal hedging portfolio under suitable (e.g. Markovian) structure conditions, and to treat in this framework the hedging of American contingent claims.

\section{Appendix A: Proofs of Selected Results}

\setcounter{equation}{0}
\setcounter{Assumption}{0}
\setcounter{Theorem}{0}
\setcounter{Proposition}{0}
\setcounter{Corollary}{0}
\setcounter{Lemma}{0}
\setcounter{Remark}{0}
\setcounter{Example}{0}

\noindent
{\bf  Proof of (\ref{ULB})-(\ref{5.10}):} $\,\,$  With $\, e_i = (0, \cdots, 0,1,0,\cdots, 0)'\,$ the $i^{th}$ unit vector in $\,\R^n\,$,
\[
\tau^{\pi}_{ii}(t)
\, = \,
\bigl( \pi(t) - e_i \bigr)' a(t)\, \bigl( \pi(t) - e_i \bigr) \,\ge\, \eps\, ||  \pi(t) - e_i ||^2
\, = \,
 \eps  \left[ \bigl( 1 -\pi_i (t) \bigr)^2 + \sum_{j \neq i} \pi_j^2 (t)\, \right] \, \ge \,  \eps    \bigl( 1 -\pi_i (t) \bigr)^2
\]
from (\ref{5.4}) and (\ref{2.3}). Back into (\ref{6.8}), this gives
\begin{eqnarray*}
\gamma^{\pi}_{*} (t)
 &\ge &\,
\frac{\eps}{2}\cdot \sum_{i=1}^n \, \pi_i(t) \,
\left[\, \bigl( 1 -\pi_i (t) \bigr)^2 \,+ \, \sum_{j \neq i} \pi_j^2 (t)\, \right]
 \\
& = & \,
\frac{\eps}{2} \cdot \left[ \, \sum_{i=1}^n  \,
\pi_i(t) \,    \bigl( 1 -\pi_i (t) \bigr)^2 \, + \,
\sum_{j=1}^n  \,
\pi_j^2(t) \,    \bigl( 1 -\pi_j (t) \bigr) \, \right]
 \\
& = & \,
\frac{\eps}{2} \cdot  \sum_{i=1}^n  \,
\pi_i(t) \,    \bigl( 1 -\pi_i (t) \bigr)  \, \ge \,
\frac{\eps}{2}\, \bigl( 1 -\pi_{(1)} (t) \bigr)\,.
\end{eqnarray*}
Similarly, we get $\,\,
\tau^{\pi}_{ii}(t)
\, \le \,
M\,  \left[\, \bigl( 1 -\pi_i (t) \bigr)^2 + \sum_{j \neq i} \pi_j^2 (t)\, \right]
\, \le \, M\,  \bigl( 1 -\pi_i (t) \bigr)\cdot$ $\,\bigl( 2 -\pi_i (t) \bigr)\,$
as claimed in (\ref{ULB}), and this leads to (\ref{5.9}) and to
\begin{eqnarray*}
\gamma^{\pi}_{*} (t)
 &\le &\,
\frac{M}{2}
\cdot  \sum_{i=1}^n  \,
\pi_i(t) \,   \bigl( 1 -\pi_i (t) \bigr)
  \, =  \,
\frac{M}{2} \cdot \left[ \,   \,
\pi_{(1)} (t) \,   \bigl( 1 -\pi_{(1)} (t) \bigr)  \, + \,
\sum_{k=2}^n  \,
\pi_{(k)} (t) \,   \bigl( 1 -\pi_{(k)} (t) \bigr) \, \right]
 \\
& \le & \,
\frac{M}{2} \cdot  \left[  \,
 \bigl( 1 -\pi_{(1)} (t) \bigr)   \, + \,
\sum_{k=2}^n  \,
\pi_{(k)} (t) \,  \right]\,=\, M\, \bigl( 1 -\pi_{(1)} (t) \bigr) \,.
\end{eqnarray*}

\smallskip
\noindent
{\bf Proof of (\ref{4.5}):} $\,\,$    Let us start by introducing the  function $\,D(x) := \Bigl(  \sum_{i=1}^n x_i^p \Bigr)^{1/p}\,$, which we shall interpret as a ``measure of diversity". An application of It\^ o's rule to the process $\, \{ D(\mu(t)),\,~ 0 \le t < \infty \}\,$ leads after some computation to the expression
\begin{equation}
\label{A.1}
\log \left( \frac{Z^{\pi^{(p)}}(T)}{Z^{\mu}(T)} \right)\,=\,
\log \left( \frac{D(\mu(T))}{D(\mu(0))} \right)\,+\, (1-p) \int_0^T \g^{\pi^{(p)}}_*(t) dt\,,~~~~ ~ 0 \le T < \infty
\end{equation}
for the value-process $\,Z_{\pi^{(p)}}(\cdot)$ of the diversity-weighted portfolio $\,\pi^{(p)} (\cdot)$ of (\ref{4.4}). Useful in the computation (\ref{A.1}) is the num\' eraire-invariance property (\ref{NI}).

Suppose that the market is weakly diverse on the finite time-horizon $\, [0,T]$, namely, that
$\,   \int_0^T \Big( 1 - \mu_{(1)}(t) \Big) \, dt \, > \, \delta \, T \,$
holds almost surely, for some $\, 0 < \delta < 1$. We have then $\,
1 =\sum_{i=1}^n \mu_i (t) \le  \sum_{i=1}^n \bigl( \mu_i (t) \bigr)^p =  \Bigl( D(\mu(t))  \Bigr)^p
 \le n^{1-p}
\,$
(minimum diversity occurs when the entire market is concentrated in one stock, and maximum diversity when all stocks have the same capitalization), so that
\begin{equation}
\label{A.3}
\log \left( \frac{D(\mu(T))}{D(\mu(0))} \right)\, \ge  \,
- \, \frac{1-p}{p}\cdot \log n\,.
\end{equation}
This provides, in particular,  the lower bound $\, Z^{\pi^{(p)}}(\cdot) /Z^{\mu}(\cdot) \ge n^{- (1-p)/p}\,$. On the other hand,   we have already remarked in section 4 that  the largest weight of the portfolio $\,\pi^{(p)} (\cdot)$ in (\ref{4.4}) does not exceed the maximum weight of the market portfolio, namely
\begin{equation}
\label{A.4}
\pi_{(1)}^{(p)}(t)\,:=\, \max_{1 \le i \le n} \pi_i^{(p)}
 (t) \,=\, \frac{\bigl(
\mu_{(1)}(t) \bigr)^p }{\sum_{k=1}^n \bigl( \mu_{(k)}(t) \bigr)^p}\, \le \, \mu_{(1)}(t)
\end{equation}
(the reverse inequality holds for the smallest weights, namely $\,
\pi_{(n)}^{(p)}(t)\,:=\, \min_{1 \le i \le n} \pi_i^{(p)}
 (t) \, \ge \,   \mu_{(n)}(t)\,$). From ({\ref{5.10}) and (\ref{A.4}) we see that the assumption (\ref{4.2})  of weak diversity  implies
\[
 \int_0^T \g^{\pi^{(p)}}_*(t) \, dt \,
\ge
\,\, \frac{\eps}{2} \cdot   \int_0^T  \Bigl( 1- \mu_{(1)}
 (t) \Bigr)  \,dt
\, >\,
\frac{\eps }{2}\cdot \delta  \, T~~~~
\]
a.s. In conjunction with (\ref{A.3}), this lead to (\ref{4.5}) via
\begin{equation}
\label{A.5}
\log \left( \frac{Z^{\pi^{(p)}}(T)}{Z^{\mu}(T)} \right)\, > \, ( 1-p) \left[ \frac{\eps T}{2} \cdot \delta   \, -\,   \frac{1}{p} \cdot \log n \right]\,.
\end{equation}

\smallskip
\noindent
{\bf Proof that the martingale property of $L(\cdot)$ (valid under condition (\ref{4.7})) proscribes  (\ref{AO}):} $\,\,$
Suppose that the exponential process $\, \{ L (t); \, 0 \le t \le T \}\,$ of (\ref{C.5}) is a $P-$martingale; then  $\, \{ \widehat W (t); \, 0 \le t \le T \}\,$ of (\ref{C.2}) is Brownian motion under the equivalent measure $\, Q_T (A) = E [ \, L(T) \cdot 1_A\,]\,$ on $\, \F(T)\,$, by the Girsanov theorem. For instance, this will be the case under the Novikov condition (4.6); cf. Theorem 3.5.1 and Proposition 3.5.12 in Karatzas \& Shreve (1991). For any extended portfolio $\, \pi( \cdot)$  we have
\[
d (Z^{z, \pi} (t) / B(t) ) \, = \, ( Z^{z, \pi} (t)  / B(t) ) \cdot \sum_{i=1}^n \sum_{\nu=1}^m  \pi_i (t)\, \sigma_{i \nu} (t) \, d \widehat W_{\nu} (t)\,,~~~~~~
Z^{z, \pi} (0) = z>0
\]
from (\ref{C.4}) and the discussion following it; this shows that the process $\, Z^{z, \pi} (\cdot) / B(\cdot)\,$ is then a martingale under $\, Q_T\,$ with moments of all orders (in particular, square-integrable). If $\, \rho(\cdot)$ is another extended portfolio, the difference $\, H(\cdot) := (Z^{z, \pi} (\cdot) - Z^{z, \rho} (\cdot)) / B(\cdot) \,$ is again a (square-integrable) martingale with $\, H(0) =0$, therefore $\, E^{Q_T} [ H(T)]=0$. But if $\, H(T) \ge 0$ holds a.s. (with respect to $P$, or equivalently with respect to $\,Q_T$), then this gives $\, H(T) =0$ a.s.  and rules out  the second requirement $\, P [ H(T) >0] >0\,$ of (\ref{AO}).

\medskip
\noindent
{\bf Proof  of (\ref{AO3}):} $\,\,$
Suppose that the  processes $\,  L(\cdot) X_i (\cdot) / B(\cdot)\,$ for $\, i = 1, \cdots , n\,$ are all martingales; then so is their sum, the process $
\,
\widetilde Z (\cdot) \,:= \, L(\cdot) Z^{\mu} (\cdot) / B(\cdot)\,$ with $\, Z^{\mu} (\cdot)  \,:=\, \sum_{i=1}^n X_i (\cdot)\,$ as in (\ref{3.1}).
With $z=1$ and $\, \vartheta^{\mu} ( t) := \sigma' (t) \mu (t)  - \vartheta  (t)\,$, the equation (\ref{C.7}) takes the form $\, d \widetilde Z^{\mu} (t) \,= \, \widetilde Z^{\mu} (t) ( \vartheta^{\mu} (t) )' \,dW(t)\,$ or equivalently
\begin{equation}
\label{11.5}
\widetilde Z^{\mu} (t) \,= \,\exp \left( \int_0^t
( \vartheta^{\mu} (s) )' \,dW(s) - \frac{1}{2}
\int_0^t || \vartheta^{\mu} (s) ||^2 \,ds
\right)\,,
\end{equation}
and we get
\[
\frac{1}{\widetilde Z^{\mu} (t)}\,=\, \exp \left( - \int_0^t
( \vartheta^{\mu} (s) )' \,d \widetilde W(s) - \frac{1}{2}
\int_0^t || \vartheta^{\mu} (s) ||^2 \,ds
\right),~~~\hbox{where}~~
\, \widetilde W (\cdot) := W (\cdot) - \int_0^{\cdot}
\vartheta^{\mu} (s)\,ds \,.
\]
Now on any given finite horizon $[0,T]$, this   process $\, \widetilde W (\cdot)\,$ is Brownian motion under the equivalent probability measure $\, \widetilde P_T (A) := E [ \widetilde Z^{\mu} (T) \cdot 1_A ] \,$ on $\, {\cal F}(T)$, and  It\^ o's rule gives
\begin{equation}
\label{11.6}
d \left( \frac{Z^{\pi} (t)}{Z^{\mu} (t)} \right) \,=\,
\left( \frac{Z^{\pi} (t)}{Z^{\mu} (t)} \right) \cdot
\sum_{i=1}^n \sum_{\nu = 1}^m
( \pi_i (t) - \mu_i (t) ) \, \sigma_{i \nu} (t)\, d
\widetilde W_{\nu} (t)
\end{equation}
for an arbitrary extended portfolio $\, \pi(\cdot)$. From (\ref{2.3}) we see that, for any such $\, \pi(\cdot)$, the ratio $\, Z^{\pi} (\cdot) /Z^{\mu} (\cdot)\,$ is a martingale under $\, \widetilde P_T \,$; in particular, $\, E^{\widetilde P_T} [\, Z^{\pi} (T) /Z^{\mu} (T) \,]=1\,$. But if $\, \pi(\cdot)$ satisfies  $\, P [\, Z^{\pi} (T)  \ge Z^{\mu} (T)\,]=1\,$, we must have  also  $\, \widetilde P_T [\,  Z^{\pi} (T) /Z^{\mu} (T)   \ge 1\,] =1\,$;   in conjunction with $\, E^{\widetilde P_T} [\,  Z^{\pi} (T) /Z^{\mu} (T)  \,]=1\,$, this leads to  $\, \widetilde P_T [\,  Z^{\pi} (T) =Z^{\mu} (T)    \,] =1\,$, or equivalently $\,   Z^{\pi} (T) =Z^{\mu} (T)   \,$  a.s. $P\,$, contradicting (4.5). Thus the process
\begin{equation}
\label{11.7}
\widetilde X_j (t) \,= \,\exp \left( \int_0^t
( \vartheta^{(j)} (s) )' \,dW(s) - \frac{1}{2}
\int_0^t || \vartheta^{(j)} (s) ||^2 \,ds
\right) \,,~~~~0 \le t < \infty
\end{equation}
of (9.8) is a strict local martingale, for some (at least one) $j \in \{ 1, \cdots, n\}$; we have set $\, \vartheta^{(k)}_{\nu} (t) := \sigma_{k \nu} (t) - \vartheta_{\nu} (t),~ \nu = 1, \cdots, n\,$, $\,$ for any $\, k \in \{ 1, \cdots, n \}$.

\smallskip
Suppose now that (\ref{AO3}) fails, i.e., that $\, \widetilde X_i (\cdot)\,$ is a martingale for some $\, i \neq j$. Then for any $T \in (0, \infty)$ the measure  $\, P^{(i)}_T (A) := E[ \widetilde X_i (T) \cdot 1_A]\,$ is a probability measure on $\F(T)$, and under this measure the process
\[
\widetilde W^{(i)} (t) \,:=\, W(t) - \int_0^t \vartheta^{(i)} (s)\,ds\,,~~~~0 \le t \le T
\]
is standard $\R^n-$valued Brownian motion. By analogy with (\ref{11.5})-(\ref{11.7}) we have now
\[
\frac{1}{\widetilde X_i (t)} \,= \,\exp \left( - \int_0^t
( \vartheta^{(i)} (s) )' \,dW^{(i)} (s) - \frac{1}{2}
\int_0^t || \vartheta^{(i)} (s) ||^2 \,ds
\right) \,,
\]
and
\[
d \left( \frac{X_j  (t)}{X_i  (t)} \right) \,=\,
\left( \frac{X_j  (t)}{X_i  (t)} \right) \cdot
\sum_{i=1}^n \sum_{\nu = 1}^m
(  \sigma_{j \nu} (t) - \sigma_{i \nu} (t)) \, d
\widetilde W^{(i)}_{\nu} (t) \,.
\]
Thus, thanks to condition (\ref{2.3}), the process  $\, X_j (\cdot) / X_i (\cdot)\,$ is a $\, P^{(i)}_T -$martingale on $[0,T]$, with moments of all orders. In particular,
\[
\frac{X_j  (0)}{X_i  (0)}\,=\, E^{P^{(i)}_T}
\left[ \, \frac{X_j  (T)}{X_i  (T)}\, \right] \,=\,
E \left[ \, \frac{L(T) X_i  (T)}{B(T) X_i  (0)} \cdot \frac{X_j  (T)}{X_i  (T)}\, \right]\,,
\]
which contradicts $\, E [ L(T) X_j (T) / B(T)] < X_j (0)\,$ and thus the strict local martingale property of $\, L(\cdot) X_j (\cdot) / B(\cdot)\,$ under $P$.

\section{Appendix B: Instantaneous Relative Arbitrage}

\setcounter{equation}{0}
\setcounter{Assumption}{0}
\setcounter{Theorem}{0}
\setcounter{Proposition}{0}
\setcounter{Corollary}{0}
\setcounter{Lemma}{0}
\setcounter{Remark}{0}
\setcounter{Example}{0}

If one is willing to dispense with the square-integrability condition (2.2) on the stock appreciation rates, then \emph{instantaneous relative arbitrage becomes possible in a diverse market:} there exists  a portfolio  $\pi(\cdot)$
with $Z^{\pi}(0) = Z^{\mu}(0)$ and $\,P[ Z^{\pi}(t) > Z^{\mu}(t),$ for all $t>0]=1$.

\smallskip
We present below a very simple, two-stock example of  such a market, in which this possibility is caused by the fact that one of the stocks is instantaneously dominating the other near $t=0$; the relative arbitrage is  created by investing all the money in the winning stock, at least for the small amount of time that it maintains the lead. The model postulates stock-prices $X_1(\cdot)$, $X_2(\cdot)$ given by
\begin{equation}
\log X_1(t) = W_1(t)\,, \quad \log X_2(t) = \Gamma(t) + W_2(t)
\end{equation}
where $(W_1,W_2)$ is a standard two-dimensional Brownian Motion and $\Gamma(t)
:= \int_0^t \gamma(u) du\,,$ for some $\gamma (\cdot)$ to de determined shortly. Of course, $X_1(0) = X_2(0) = 1$.

For such a market the diversity condition (6.1) becomes $\, P [ \delta <
\mu_1(t) < 1 - \delta$, for all $t \ge 0]=1\,$, for some $\, \delta \in (0,1/2)$.
Setting $Y(t) := \log X_2(t) - \log X_1(t)\,$,
diversity is equivalent to $\, -\eta < Y(t) < \eta\,$, where
$\,\eta :=  \log (1-\delta)- \log \delta  > 0$. Note also that $Y(\cdot) = \Gamma(\cdot) - W(\cdot)$, where
$W (\cdot) := W_1 (\cdot) - W_2 (\cdot)\,$ is a (scaled) Brownian Motion. Now consider another number $\delta'$ satisfying $0 < \delta <
\delta' < 1/2$, introduce $\,\eta' :=  \log (1-\delta')- \log \delta'  < \eta$ as above, and define the stopping time $\,T_1 := \inf \{\,t > 0 \, | \, Y(t) \notin (-\eta', \eta')\,\} >0$. For any given $\,\alpha \in (0,1/2)\,$, let
\[
\gamma(t):=
\left\{
\begin{array}{ll}
    \alpha t^{\alpha-1}, & \hbox{$t \leq T_1$} \\
    q(Y(t)), & \hbox{$t > T_1$} \\
\end{array}
\right\},
\]
where $q(\cdot)$ is a function on $(-\eta, \eta)$ with such
singularities at $\,-\eta$ (infinite push-up) and $\eta$
(infinite pull-down) as to keep the process $Y(\cdot)$ inside these
bounds; for instance, one can use  log-pole type singularities as in Theorem 6.1. Thus the market of (12.1) is diverse.

Clearly $\Gamma(t) = t^\alpha$ on $[0,T_1]$, and the Law of
the Iterated Logarithm implies     $\, \lim_{t \downarrow 0}
W(t)/\Gamma(t) = 0$, so  $\,Y(t) = \Gamma(t) - W(t) = \Gamma(t) \cdot [1 - (W(t)/\Gamma(t))]\,$ strictly dominates $\, \Gamma(t)/2$ near
$\, t =0$. This means that, if we consider $T_2 := \inf \{\,t > 0 \, | \, Y(t) = \Gamma(t)/2\,\}$, then $\,T_2 > 0$ and $Y(t) \geq \Gamma(t)/2$ on
$[0,T_2]$, i.e., $\,Y(t) > 0$ and  $\,X_2(t) > X_1(t)$ hold on $(0, T_2]$. Finally, we define the portfolio
\[
\pi(t) := \left\{
\begin{array}{ll}
    (0,1)^{\prime}, & \hbox{$t \leq T_2$} \\
    \mu(t), & \hbox{$t > T_2$} \\
\end{array}
\right\}
\]
which {\sl invests everything on the second stock until time $T_2\,$,
then switches to the market portfolio}. In the interval $\,[0,T_2]$ we have $\,Z^{\pi} (t) - Z^{\mu} (t) = X_2(t) - X_1(t)$: this is because we start with $X_1(0)=X_2(0)=1$, so that $\, Z^{\pi}
(0) = Z^{\mu} (0) = 2$ and  $Z^{\pi} (t) = 2
X_2(t)\,$, $\,Z^{\mu} (t) = X_1(t) + X_2(t)$ hold on $\,[0,T_2]$. After time $T_2$ the
two portfolios have the exact same performance. We just saw that
$\,X_2(t) > X_1(t)$ for all $t \in (0,T_2]$, so this will imply
$\,Z^{\pi} (t) > Z^{\mu} (t)$ for all $t > 0$, as claimed.

Of course, in order to create this possibility we had to use a rate of growth $\,\gamma (\cdot) $
such that $\, ||\gamma (\cdot) ||^2 $ (as well as $\,||b (\cdot) ||^2
   $, $\, ||\vartheta (\cdot) ||^2 $)
is not locally integrable: near zero, we have $\,||\gamma (\cdot) ||^2  \sim c
t^{-\beta}$ with $\beta > 1$. As a result,  no ``local martingale density'' process $\,L(\cdot)$ as in (9.5) can be constructed for the model of (12.1). It is  easy to see that the existence of such a process $\,L (\cdot)$ proscribes this kind of instantaneous relative arbitrage.

\smallskip
\noindent
{\it 12.1 Remark:}~ This example is reminiscent of similar ones in Levental \& Skorohod (1995). These authors also show the following result (Corollary 2 on p.909): Suppose that one insists on the square-integrability condition (2.2) and assumes that $\, m=n$, that  $ \sigma (\cdot)$ is invertible, and that the filtration ${\bf F}$ is generated by the driving Brownian Motion $W(\cdot)$. Then there exists a trading strategy $\f (\cdot)$ with $\, P [ \, Z^{0,\f} (T) \ge 0\,] = 1\,$, $\, P [ \, Z^{0,\f} (T) > 0\,] >0\,$ and $\, P [ \, Z^{0,\f} (t) \ge q_{\f, T} \,, \forall \, 0 \le t \le T \,] = 1\,$ for some $\, T \in (0,\infty)$ and $q_{\f, T} \in \R$ (``arbitrage with tame trading strategies"), if and only if there does not exist on $\, {\cal F}(T)$ any  probability measure equivalent to $P$   under which  the process $\, \widehat{W} (t),\, 0 \le t \le T\,$ of (9.2) becomes Brownian Motion.

As we saw in Remark 9.1, such a measure indeed fails to exist if the market $\, {\cal M}$ is weakly diverse on $\, [0,T]$. In other words, {\sl a market with (2.2),} $\, m=n$, {\sl Brownian filtration and $ \sigma (\cdot)$ invertible, contains ``arbitrage with tame trading strategies" over} any {\sl finite horizon $[0,T]$ on which it is weakly diverse}.


\section{Acknowledgements}

\medskip
\noindent
We are grateful for the helpful remarks offered by seminar audiences at the Sloan School, MIT, at the Mathematical Institute in Oberwolfach, and at the University of Athens, particularly by Professors Julien Hugonnier, Leonid Kogan, Ralf Korn, Mark Lowenstein, Andrew Lyasoff, Evangelos Mageirou, Steven Shreve, Wolfgang Stummer, Dimitri  Vayanos and Jiang Wang. We are also indebted to Dr. Adrian Banner  for a number of discussions that helped sharpen  our thinking about these problems.

A significant part of this work was completed in the spring semester of 2002, while the second author was on sabbatical leave at the Cowles Foundation for Research in Economics, Yale University. He is grateful to the Foundation for its hospitality.

\bigskip
\section{References}

\medskip
\noindent DELBAEN, F. \& SCHACHERMAYER, W. (1995) Arbitrage possibilities in Bessel processes and their relations to local martingales. {\it Probability Theory \& Related Fields}~ {\bf 102}, 357-366.

\medskip
\noindent DELBAEN, F. \& SCHACHERMAYER, W. (1998)
The Fundamental Theorem of Asset Pricing for Unbounded Stochastic Processes. {\it
Mathematische Annalen}~ {\bf 312},  215-250.

\medskip
\noindent
ELWORTHY, K.D., X.-M. LI \& M. YOR (1997)~  On the tails of the supremum and the quadratic variation of strictly local martingales. In {\sl S\' eminaire de Probabilit\' es XXXI}, {\it Lecture Notes in Mathematics}  {\bf 1655}, pp. 113-125.  Springer-Verlag, Berlin.

\smallskip
\noindent FERNHOLZ, E.R.  (1999)  On the diversity of equity markets. {\it Journal of Mathematical Economics}
~{\bf 31}, 393-417.

\smallskip
\noindent FERNHOLZ, E.R.  (2001)  Equity portfolios generated by ranked market-weights. {\it Finance \& Stochastics}
~{\bf 5}, 469-486.

\medskip
\noindent FERNHOLZ, E.R.  (2002) {\it  Stochastic Portfolio Theory.}  Springer-Verlag, New York.

\medskip
\noindent KARATZAS, I., LEHOCZKY, J.P.,  SHREVE, S.E.  \& XU, G.L.  (1991)  Martingale and duality methods for utility maximization in an incomplete market. {\it SIAM Journal on Control \& Optimization} {\bf 29}, 702-730.

\medskip
\noindent KARATZAS, I. \&  SHREVE, S.E.  (1991)  {\it Brownian Motion and Stochastic Calculus}.  Second Edition. Springer-Verlag, New York.

\medskip
\noindent KARATZAS, I.  \& SHREVE, S.E.  (1998)  {\it Methods of Mathematical Finance.}  Springer-Verlag, New York.

\medskip
\noindent LEVENTAL, S. \& SKOROHOD, A. (1995) A necessary and sufficient condition for absence of arbitrage with tame portfolios.  {\it Annals of Applied Probability}~ {\bf 5}, 906-925.

\medskip
\noindent LOEWENSTEIN, M. \& WILLARD, G.A. (1995) Local martingales, arbitrage and viability.  {\it Economic Theory}~ {\bf 16}, 135-161.

\medskip
\noindent SAMUELSON, P.A. (1965)  Proof that properly anticipated prices fluctuate randomly. {\it Industrial Management Review} ~{\bf 6}, 41-50.

\medskip
\noindent STUMMER, W. (1993) The Novikov and entropy conditions for multi-dimensional diffusion processes with singular drift. {\it Probability Theory \& Related Fields}~ {\bf 97}, 515-542.

\medskip
\noindent VERETENNIKOV, A. Yu. (1981)  On strong solutions and explicit formulas for solutions of stochastic integral equations. {\it Math. USSR Sbornik} ~{\bf 39}, 387-403.

\end{document}